\documentclass[twoside,twocolumn,9pt]{article}
\usepackage{extsizes}
\usepackage[super,sort&compress,comma]{natbib} 
\usepackage[version=3]{mhchem}
\usepackage[left=1.5cm, right=1.5cm, top=1.785cm, bottom=2.0cm]{geometry}
\usepackage{balance}
\usepackage{mathptmx}
\usepackage{sectsty}
\usepackage{graphicx} 
\usepackage{lastpage}
\usepackage[format=plain,justification=justified,singlelinecheck=false,font={stretch=1.125,small,sf},labelfont=bf,labelsep=space]{caption}
\usepackage{float}
\usepackage{fancyhdr}
\usepackage{fnpos}
\usepackage[english]{babel}
\addto{\captionsenglish}{%
  \renewcommand{\refname}{Notes and references}
}
\usepackage{array}
\usepackage{droidsans}
\usepackage{charter}
\usepackage[T1]{fontenc}
\usepackage[usenames,dvipsnames]{xcolor}
\usepackage{setspace}
\usepackage[compact]{titlesec}
\usepackage{hyperref}
\usepackage{pdfpages} % include pdfs
\usepackage{pgffor} % for loops

\makeatletter
\AtBeginDocument{\let\LS@rot\@undefined}
\makeatother

\def\supplementfilename{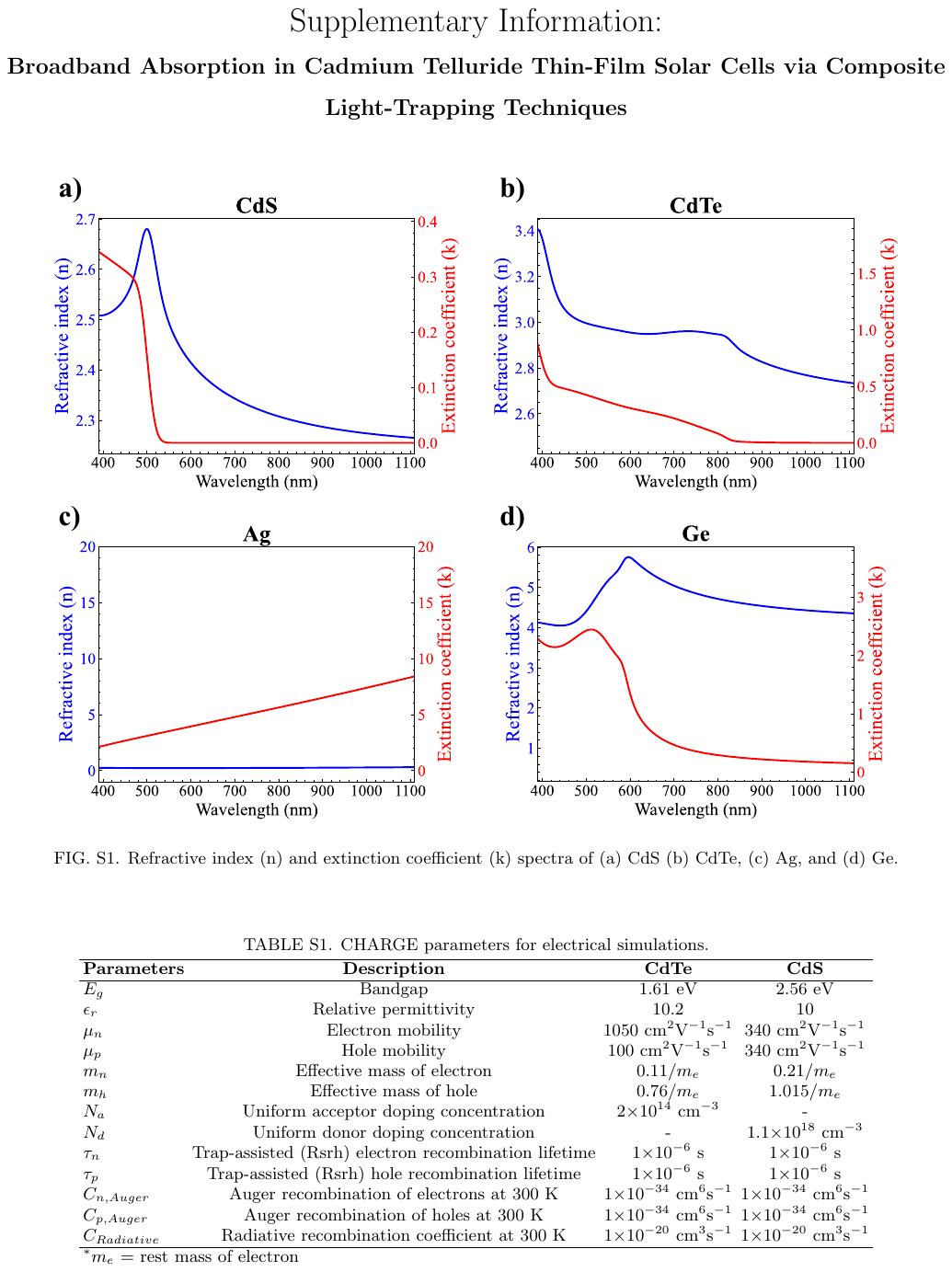}

\pdfximage{\supplementfilename}
\def\numbersupplementpages{\the\pdflastximagepages}

\newif\ifarXiv
\arXivtrue 
\newcolumntype{h}{>{\centering\let\newline\\\arraybackslash\hspace{0pt}}m}

\definecolor{cream}{RGB}{222,217,201}

\begin{document}

\pagestyle{fancy}
\thispagestyle{plain}
\fancypagestyle{plain}{
%%%HEADER%%%
\renewcommand{\headrulewidth}{0pt}
}
%%%END OF HEADER%%%

%%%PAGE SETUP - Please do not change any commands within this section%%%
\makeFNbottom
\makeatletter
\renewcommand\LARGE{\@setfontsize\LARGE{15pt}{17}}
\renewcommand\Large{\@setfontsize\Large{12pt}{14}}
\renewcommand\large{\@setfontsize\large{10pt}{12}}
\renewcommand\footnotesize{\@setfontsize\footnotesize{7pt}{10}}
\makeatother

\renewcommand{\thefootnote}{\fnsymbol{footnote}}
\renewcommand\footnoterule{\vspace*{1pt}% 
\color{cream}\hrule width 3.5in height 0.4pt \color{black}\vspace*{5pt}} 
\setcounter{secnumdepth}{5}

\makeatletter 
\renewcommand\@biblabel[1]{#1}            
\renewcommand\@makefntext[1]% 
{\noindent\makebox[0pt][r]{\@thefnmark\,}#1}
\makeatother 
\renewcommand{\figurename}{\small{Fig.}~}
\sectionfont{\sffamily\Large}
\subsectionfont{\normalsize}
\subsubsectionfont{\bf}
\setstretch{1.125} %In particular, please do not alter this line.
\setlength{\skip\footins}{0.8cm}
\setlength{\footnotesep}{0.25cm}
\setlength{\jot}{10pt}
\titlespacing*{\section}{0pt}{4pt}{4pt}
\titlespacing*{\subsection}{0pt}{15pt}{1pt}
%%%END OF PAGE SETUP%%%

%%%FOOTER%%%
\fancyhead{}
\renewcommand{\headrulewidth}{0pt} 
\renewcommand{\footrulewidth}{0pt}
\setlength{\arrayrulewidth}{1pt}
\setlength{\columnsep}{6.5mm}
\setlength\bibsep{1pt}
%%%END OF FOOTER%%%

%%%FIGURE SETUP - please do not change any commands within this section%%%
\makeatletter 
\newlength{\figrulesep} 
\setlength{\figrulesep}{0.5\textfloatsep} 

\newcommand{\topfigrule}{\vspace*{-1pt}% 
\noindent{\color{cream}\rule[-\figrulesep]{\columnwidth}{1.5pt}} }

\newcommand{\botfigrule}{\vspace*{-2pt}% 
\noindent{\color{cream}\rule[\figrulesep]{\columnwidth}{1.5pt}} }

\newcommand{\dblfigrule}{\vspace*{-1pt}% 
\noindent{\color{cream}\rule[-\figrulesep]{\textwidth}{1.5pt}} }

\makeatother
%%%END OF FIGURE SETUP%%%

%%%TITLE, AUTHORS AND ABSTRACT%%%
\twocolumn[
  \begin{@twocolumnfalse}
\sffamily
\begin{tabular}{p{0.96\textwidth}}

\centering\LARGE{\textbf{Broadband Absorption in Cadmium Telluride Thin-Film Solar Cells via Composite Light Trapping Techniques$^\dag$}} \\%Article title goes here instead of the text "This is the title"
\vspace{0.3cm}

\large{Asif Al Suny,\textit{$^{a}$} Tazrian Noor,\textit{$^{a}$} Md. Hasibul Hossain,\textit{$^{a}$} A. F. M. Afnan Uzzaman Sheikh,\textit{$^{a}$} Mustafa Habib Chowdhury$^{\ast}$\textit{$^{a}$}} \\%Author names go here instead of "Full name", etc.

\end{tabular}

\begin{tabular}
{m{0.06\textwidth}p{0.8\textwidth}m{0.06\textwidth}}

 & \noindent\normalsize{Composite light-trapping structures offer a promising approach to achieving broadband absorption and high efficiency in thin-film solar cells (TFSCs) in order to accelerate sustainable energy solutions. As the leading material in thin-film solar technology, cadmium telluride (CdTe) faces challenges from surface reflective losses across the solar spectrum and weak absorption in the near-infrared (NIR) range. This computational study addresses these limitations by employing a dual light trapping technique: the top surfaces of both the CdS and CdTe layers are tapered as nanocones (NCs), while germanium (Ge) spherical nanoparticles (NPs) are embedded within the CdTe absorber layer to enhance broadband absorption. Numerical simulations using Finite-Difference Time Domain (FDTD) and other methods are used to optimize the parameters and configurations of both nanostructures, aiming to achieve peak optoelectronic performance. The results show that a short-circuit current density (J\textsubscript{sc}) of 35.38 mA/cm\textsuperscript{2} and a power conversion efficiency (PCE) of 27.76\% can be achieved with optimal nanocone (NC) texturing and spherical Ge nanoparticle (NP) configurations, a 45.45\% and 80.72\% increase compared to baseline structure in J\textsubscript{sc} and PCE respectively. To understand the enhancement mechanisms, the study includes analyses using diffraction grating theory and Mie theory. Fabricability of these structures is also evaluated. Furthermore, an additional study on the effects of incident angle variation and polarization change demonstrates that the optimal structure is robust under practical conditions, maintaining consistent performance.} \\ & %The abstrast goes here instead of the text "The abstract should be..."
\end{tabular}

 \end{@twocolumnfalse} \vspace{0.6cm}

  ]
%%%END OF TITLE, AUTHORS AND ABSTRACT%%%

%%%FONT SETUP - please do not change any commands within this section
\renewcommand*\rmdefault{bch}\normalfont\upshape
\rmfamily
\section*{}
\vspace{-1cm}

%%%FOOTNOTES%%%

\footnotetext{\textit{$^{a}$~Department of Electrical and Electronic Engineering, Independent University, Bangladesh, Plot 16, Block B, Aftabuddin Ahmed Road, Bashundhara Residential Area, Dhaka 1229, Bangladesh.}}
\footnotetext{\textit{$^{\ast}$~E-mail: mchowdhury@iub.edu.bd}}

%Please use \dag to cite the ESI in the main text of the article.
%If you article does not have ESI please remove the the \dag symbol from the title and the footnotetext below.
\footnotetext{\dag~Supplementary Information available.}
%additional addresses can be cited as above using the lower-case letters, c, d, e... If all authors are from the same address, no letter is required

%%%END OF FOOTNOTES%%%

%%%MAIN TEXT%%%%
\section{Introduction}

Sustainable energy development is one of the major aspects of achieving sustainability which focuses on generating clean energy without emitting greenhouse gases. Solar cells play a vital role in achieving sustainability goals and will continue to do so in the upcoming decades \cite{SolEneTec}. Solar cells use earth-abundant solar energy to convert it into electrical energy, the most useful form of energy. Presently fossil fuels dominate global electricity generation, yet renewable energy sources are rapidly closing the gap \cite{IEA-Ren}. Although less than 5\% of the present global electricity generation contribution comes from solar photovoltaics, and this falls short of other forms of renewable source like wind and hydropower, solar cells are expected to soon overtake the latter two and contribute to 16\% by the year 2030 \cite{IEA-Ren}. By that time, the total solar manufacturing capacity is expected to exceed 1200 GW, i.e., almost double the present capacity \cite{IEA-Wor}. Thin-film solar cells (TFSCs) are more sustainable compared to the market-dominant crystalline silicon-based solar cells due to the fact that they require less material (being almost 100 times thinner), leave reduced carbon footprint, and have a lower fabrication cost \cite{IntPhoAlt}. CdTe TFSCs dominate the commercial thin-film production and and is expected to have a market worth of 27.11 billion dollars by 2030 \cite{EleOptPara}. Being a direct bandgap material and having an optimal bandgap of 1.45 eV, CdTe-based TFSCs have the potential to achieve the maximum power conversion efficiency (PCE) possible by the Shockley-Queisser limit \cite{TheHisPho}. Its high absorption coefficient allows it to absorb almost all the photons of the solar AM 1.5G solar spectrum, having only a 2 $\mu$m thick absorber layer \cite{HigImpLig}. Even at this thickness, the CdTe/CdS II-VI TFSCs are quite flexible, making them a potential choice to be integrated into portable electronic devices, windows, roofs, facades, electric cars, or even large-scale commercial purposes \cite{HigImpLig}. From sustainability prospect, the production of CdTe emits 6 times lower carbon dioxide than Si with high-throughput manufacturing methods \cite{US-DOE}. Furthermore, once it reaches the end of its lifetime, 90\% of CdTe modules can be recycled \cite{US-DOE}. Despite having potential, high surface reflective loss over all wavelengths and poor light absorption in near-infrared (NIR) regions limit the performance of CdTe TFSCs.

Light trapping provides a meaningful solution to this problem and can be done via integrating nanostructures like nanoparticles (NPs), nanowires, nanogratings, photonic crystals, etc. \cite{ResProPlas, MulNanMat, PhoCryStr,STI_CNT}. Surface texturing is another aspect that has been widely used to prevent surface reflective losses \cite{SurTexMet,STI_ST}. However, any light trapping technique with a certain configuration can only provide performance elevation over a limited wavelength range. Broadband absorption, i.e., absorption over a broad wavelength range, can only be achieved by employing composite light trapping techniques. Such composite light trapping works in turns to a specific wavelength to achieve broadband absorption. 

Previous studies have shown that by utilizing composite light trapping techniques, absorption over a broad wavelength can be significantly increased while achieving unique properties such as insensitivity to polarization and incident angle change, homogenization of optical fields, etc \cite{HomOptFie, OptThinSi}. In a theoretical study, H. Li et al. have shown composite light trapping consisting of SiO\textsubscript{2} nanospheres on the front surface and Ag hemispheres on the rear surface can achieve absorption enhancement in both short and long wavelengths with only a 100 nm thick a-Si absorber layer \cite{TheInvBro}. In another study, Pritom et al. have proposed a parabolic surface texture with gold (Au) NPs on top which can provide broadband absorption between the wavelength range of 300 nm to 1600 nm \cite{PlaEnhPara}. Furthermore, a similar parabolic texture but embedded dielectric NP can offer a 52.3\% absorption gain compared to the baseline structure \cite{EnhLigAbsThin}. On the other hand, Y. Yin et al. were able to maintain high absorption efficiency over an incident angle variation of 0$^{\circ}$ to 60$^{\circ}$ by combining semicircular dielectric grooves at the front and trapezoidal metal reflector at the back of a 100 nm thick a-Si TFSCs \cite{DesPlasSol}. All the mentioned works are mainly based on either a-Si or perovskite solar cells. However, in terms of CdTe TFSCs, such composite light trapping methods are still unexplored despite the potential to significantly improve absorption and other performance parameters.

In this aspect, this paper presents a composite light trapping technique that consists of a nanocone (NC) shaped surface texture of both the CdS-CdTe surface layers and embedding Germanium (Ge) spherical NP into the CdTe absorber layer. The idea is to suppress surface reflection over all the wavelengths, especially in the CdTe absorption window (400 nm to 800 nm) in order to maximize the short-circuit current density (J\textsubscript{sc}). Additionally, the embedded Ge NP would facilitate necessary light trapping to increase absorption for the rest of the wavelengths (800 nm to 1100 nm). Compared to other texture shapes, cone-shaped texture has more desirable performance in suppressing surface reflection for a broad wavelength range and also shows insensitivity to incident and polarization angle variation \cite{BroOmnAnt, NanPhoMan}. Prashant et al. recently demonstrated NC structured GaAs solar cells have achieved an average absorption of 94\%, outperforming their counterpart nanowire-shaped solar cells which show an average absorption of 80.8\% with the angle of incidence between 0 to 70 degrees. Additionally, NC texture can also function as a diffraction grating to trap light into diffraction modes to achieve enhanced light absorption \cite{OptNanPara}. To realize the highest diffraction efficiency, careful optimization of the cone base diameter and height is important \cite{OptNanPara}. Nevertheless, Ge NP is considered to be an environment-friendly green material due to its non-toxic nature \cite{OptProGer}. As found in the literature, high-index semiconductor NPs can open doors to achieve super absorbing, highly efficient solar cells with unique properties of broadband absorption, insensitivity to polarized, oblique-angled light, and different environmental conditions \cite{LigManPho, EnhPhoPer}.  Semiconductor NPs with a high dielectric constant are superior to metallic NPs in terms of ultra-low light-to-heat conversion and simultaneous electric and magnetic resonance excitation \cite{DirScaGer}. At high temperatures, noble metal NP suffers poor thermal and chemical stability \cite{AltPlasMat, LowPlasMet}. On the contrary, semiconductor NPs are less prone to the deterioration of absorption and scattering efficiency at high temperatures \cite{TheResSem}. Ge being a semi-metal NP is no different and can be considered a viable alternative to plasmonic metal NP due to its high refractive index \cite{DirScaGer, ResOptAbs, DirLigSca}. Along with a high refractive index, Ge possesses some interesting properties such as high charge carrier mobilities ($\mu$\textsubscript{e} = 3900 cm\textsuperscript{2}V\textsuperscript{-1}s\textsuperscript{-1}, $\mu$\textsubscript{h} = 1900 cm\textsuperscript{2}V\textsuperscript{-1}s\textsuperscript{-1}) and narrow bandgap (0.67 eV at 300 K) \cite{RecAdvGer}. Ge is among a few materials whose Re[$\epsilon$] exceeds 20, even Silicon’s (Re[$\epsilon$] = 11.70) \cite{HanOptCon}. Such a high positive permittivity value allows Ge to meet the Mie resonance condition in the subwavelength regime and can have comparable strength to the plasmon resonance of metallic NPs \cite{ResOptAbs}. To harness the light scattering capacity of Ge NP to the fullest and utilize it to increase CdTe TFSCs efficiency, optimization of its size and placement is essential.

\section{Methodology}
\subsection{Simulation setup}

\begin{figure*}
 \centering
 \includegraphics[width=\textwidth]{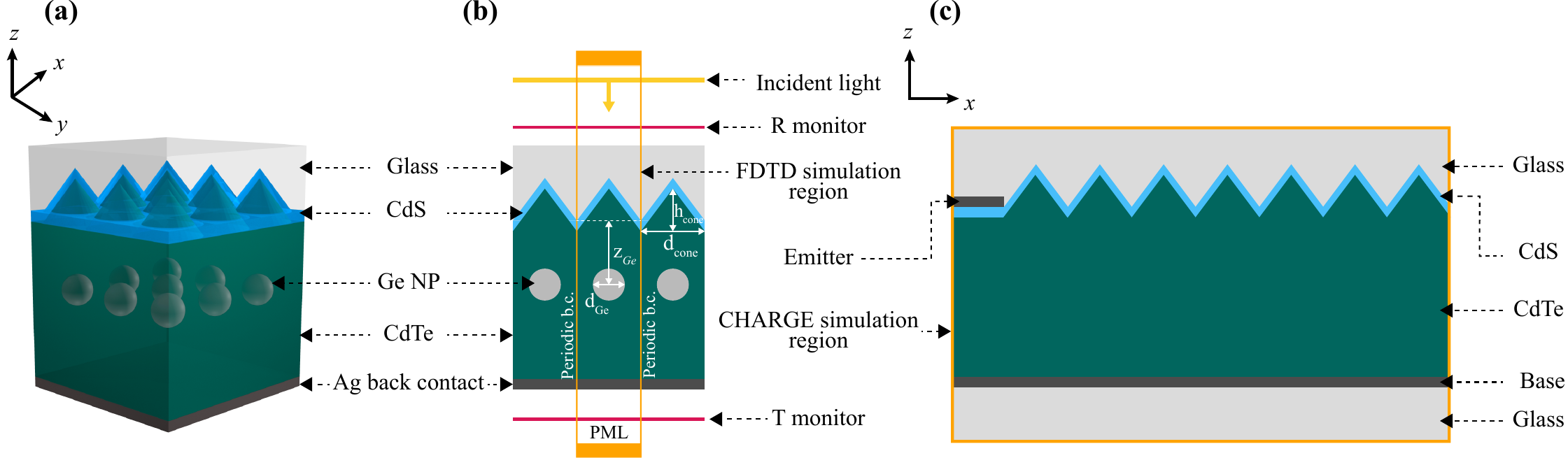}
 \caption{(a) 3D schematic view of the optimal structure (with NC texture \& Ge NP), (b) FDTD
simulation setup for optical analysis, (c) CHARGE simulation setup for electrical analysis.}
 \label{fdtd_charge_setup}
\end{figure*}

Fig. \ref{fdtd_charge_setup}(a) shows a 3D view of the proposed NC texture and embedded spherical Ge NP in a 3×3 array. Using the commercially available ANSYS-Lumerical software package FDTD and CHARGE suite, simulations have been carried out to find the optimal configuration for NC texture and embedded spherical Ge NP that results in maximum absorption and optimal performance. Overall, the simulations can be divided into two parts - optical simulations and electrical simulations. Using discrete spatial and temporal grid cells (Yee cells), the FDTD solver solves Maxwell’s equations to calculate the charge carrier generation, absorption, reflection, transmission, and electric fields \cite{AbsEnhGaAs}. For optical simulations, wavelength-dependent material data for CdS, CdTe, Ge, and Ag are all taken from established sources \cite{OptDesFab, HanOptCon, OptConFar} and shown in Fig. S1. Afterwards, the carrier generation data from FDTD are imported into the CHARGE solver to calculate electrical parameters i.e. short-circuit current density (J\textsubscript{sc}), open-circuit voltage (V\textsubscript{oc}), fill factor (FF) and power conversion efficiency (PCE) by solving Poisson’s and drift-diffusion equations. 

Fig. \ref{fdtd_charge_setup}(b) illustrates the detailed FDTD simulation setup for 3D optical simulation. Coming from the top, glass, CdS, CdTe, and Ag are all stacked together to form the whole structure illuminated by a plane wave source of AM1.5G solar irradiance (1000 W/m\textsuperscript{2}) along the z-axis. The source wavelength range has been taken from 400 nm to 1100 nm. Considering the two-dimensional symmetry of both the NC texture and spherical Ge NP, antisymmetric and symmetric boundary conditions are imposed in the x and y direction to infinitely mirror the unit cell. The unit cell consists of only one pair of CdS-CdTe NC-shaped textures and a single Ge NP. On the contrary, to mimic the real surrounding environment, perfectly matched layer (PML) boundary conditions are applied in both of the z directions where any light that either passes through the solar cell or is reflected from the solar cell will be absorbed by the PML layers. Here the CdS window layer and CdTe absorber layer thickness of the baseline structure (without texture) are taken at 100 nm and 1500 nm, respectively \cite{OptAbsEnhCdTe}. The 100 nm Ag metal contact is working here as a back reflector at the bottom. Two frequency domain power monitors are used to record the amount of reflected and transmitted light, placed one above the CdS layer (R monitor that measures the power of the reflected light) and the other below the Ag back contact (T monitor that measures the power of the transmitted light). The simulation will shut off when the total energy of the simulation volume reaches 10\textsuperscript{-5} of the incident maximum energy. The polarization and incident angle are both kept at 0$^{\circ}$ during simulations of every study in this work, except for the polarization angle and incident angle variation studies. To calculate the electric field distribution of the absorber layer, another frequency domain power monitor was placed along the z-axis (not shown in the figure). There are a total of four parameters that are being optimized in this study. NC-shaped texture’s base diameter, d\textsubscript{cone} and height, h\textsubscript{cone} were optimized first. It is worth mentioning that no space was considered between the two successive NCs, meaning the period of NC is the same as its base diameter. Later upon obtaining the optimal texture structure, Ge NP diameter, d\textsubscript{Ge}, and position, i.e., depth from the top surface of the flat CdS layer, z\textsubscript{Ge} were optimized.

Fig. \ref{fdtd_charge_setup}(c) illustrates the CHARGE simulation setup for 2D electrical simulation. Although the optical simulations were conducted for only a single period, the generation data was then averaged and unfolded for multiple periods for the electrical simulations. To successfully collect the charge carriers, an emitter, and a base metal contact were placed at the top and bottom, respectively. No generation data were considered under the top emitter metal contact to account for the shadowing effect. The \textit{p}-type CdTe layer and \textit{n}-type CdS layer were uniformly doped at a concentration of 2×10\textsuperscript{14} cm\textsuperscript{-3} and 1.1×10\textsuperscript{18} cm\textsuperscript{-3}, respectively \cite{ChaModCdSCdTe}. Finally, to obtain the P-V and J-V characteristics curve of the solar cells, a voltage sweep from 0 V to 1.2 V with 0.05 V step size was considered to run the simulation. Fill factor and PCE were also calculated from the P-V and J-V curves. The CHARGE parameters for the electrical simulations are tabulated in Table S1. Lumerical scripting language and Python were used to extract, analyze, and plot the data collected from the optical and electrical simulations.

\subsection{Numerical analysis}

The finite-difference time-domain (FDTD) method, used to perform the optical simulations of the design, solves Maxwell’s equations without approximating any physical components. The generation rate, absorption plot, and electric field plot are generated using FDTD method, as described in the later parts of this study.

\begin{figure*}[!t]
 \centering
 \includegraphics[width=\textwidth]{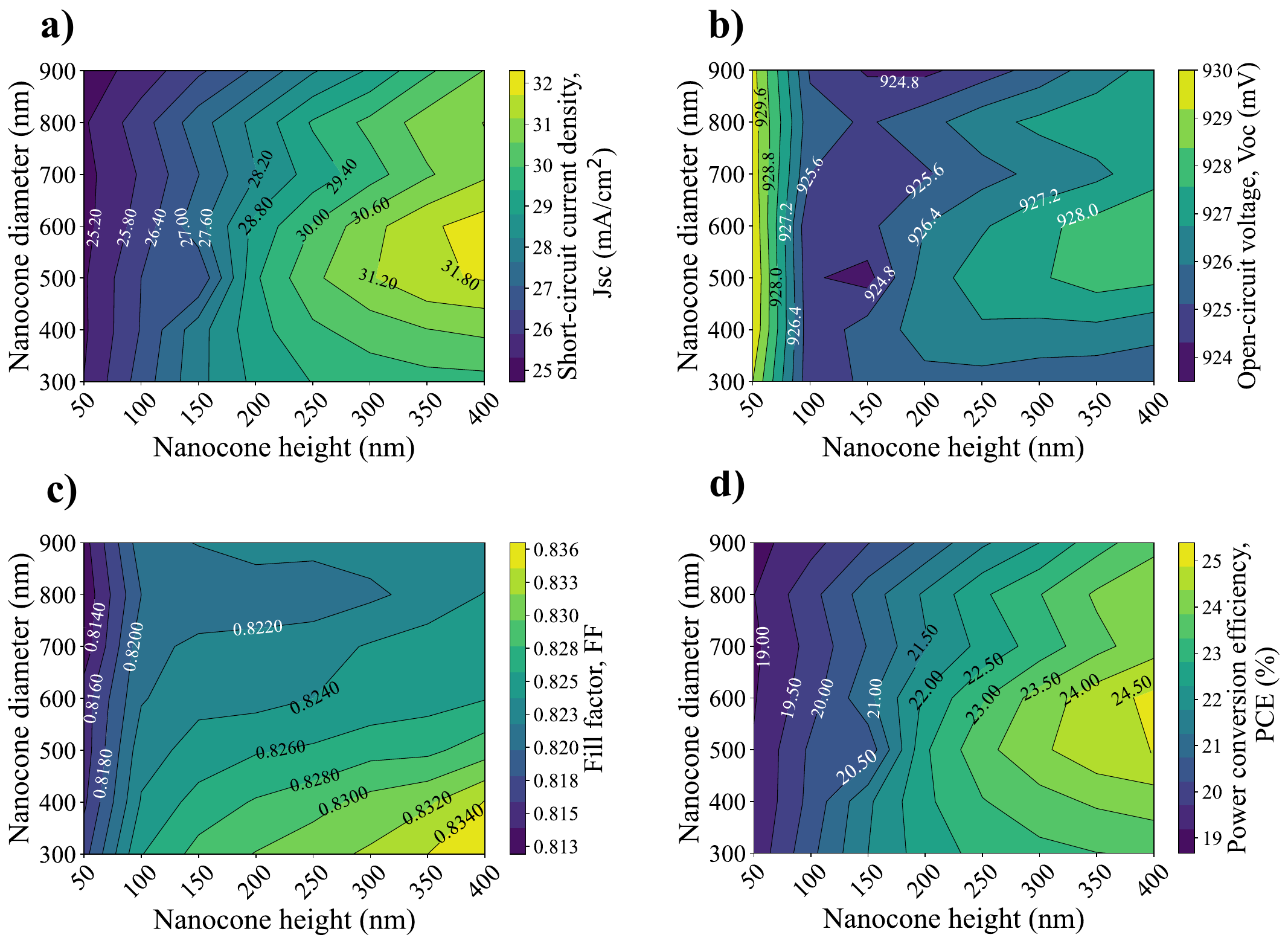}
 \caption{Contour plot of photovoltaic parameters, i.e., (a) short-circuit current density (J\textsubscript{sc}), (b)
open-circuit voltage (V\textsubscript{oc}), (c) fill factor (FF), and (d) power conversion efficiency (PCE), as a function of NC texture height (h\textsubscript{cone}) and base diameter (d\textsubscript{cone}). J\textsubscript{sc}, V\textsubscript{oc}, fill factor, and PCE for bare CdTe TFSCs are 24.33 mA/cm\textsuperscript{2}, 978 mV, 0.6455, and 15.36\%, respectively.}
 \label{texture_optimization}
\end{figure*}

Depending on its property, a material can transmit, reflect, and absorb a certain percentage of photons that strikes it. In the context of a solar cell, we need the substrate to absorb as much of the incident light as possible, so that more of the photons are used to generate electron-hole pairs for current generation. Thus, the relative absorbance is an important factor for the solar cell light absorbing material. The absorbance is calculated using the following formula \cite{absorbance},
\begin{equation}
\label{absorbance}
A(\lambda)=1-R(\lambda)-T(\lambda)
\end{equation}
where A($\lambda$) is the absorbance, R($\lambda$) is the reflectance, and T($\lambda$) is the transmittance as a function of wavelength ($\lambda$) of the incident light for the material.

The optical power absorption per unit volume (P\textsubscript{abs}) is calculated as follows \cite{num_analysis},
\begin{equation}
P_{\text{abs}}(\mathbf{r}, \omega) = -0.5 \left| \mathbf{E}(\mathbf{r}, \omega) \right|^2 \text{Im}[\epsilon(\mathbf{r}, \omega)]
\end{equation}
which in turn is used to derive the following expression for the generation rate G(\textbf{r}) of the light source in FDTD \cite{num_analysis},
\begin{equation}
G(\mathbf{r}) = \frac{P_{\text{abs}}(\mathbf{r}, \omega)}{h\omega} = -0.5 \frac{\left| \mathbf{E}(\mathbf{r}, \omega) \right|^2 \text{Im}[\epsilon(\mathbf{r}, \omega)]}{h}
\end{equation}
where $\omega$ is the angular frequency, \textbf{E} is the complex electric field, and Im($\epsilon$) is the imaginary part of the dielectric constant associated with loss.

The generation rate is exported in CHARGE to calculate the electrical parameters, such as short-circuit current density (J\textsubscript{sc}), open-circuit voltage (V\textsubscript{oc}), power conversion efficiency (PCE), and fill factor. The quantum efficiency of a solar cell is the ratio of the power of the light absorbed by solar cell (P\textsubscript{abs}($\lambda$)) to the power of the light incident (P\textsubscript{in}($\lambda$)) on it, as given in the equation below \cite{ansys},
\begin{equation}
QE(\lambda) = \frac{P_{\text{abs}}(\lambda)}{P_{\text{in}}(\lambda)}
\end{equation}
J\textsubscript{sc} is the maximum amount of current generated per unit area by the solar cell under short-circuit condition, i.e., the voltage is 0, when light incidents on the solar cell. Each photon is depicted to generate an electron ideally, and the following formula is used to calculate J\textsubscript{sc} \cite{num_analysis},
\begin{equation}
\label{jsc}
J\textsubscript{sc}=e\int\frac{\lambda}{hc}QE(\lambda)I\textsubscript{AM1.5G}(\lambda)d\lambda
\end{equation}
where \textit{e} is the charge of an electron, $\lambda$ is the wavelength, \textit{h} is the plank’s constant, \textit{c} is the speed of light, and I\textsubscript{AM1.5G}($\lambda$) is the spectral irradiance of incident light.

\begin{figure*}[!t]
 \centering
 \includegraphics[height=7cm]{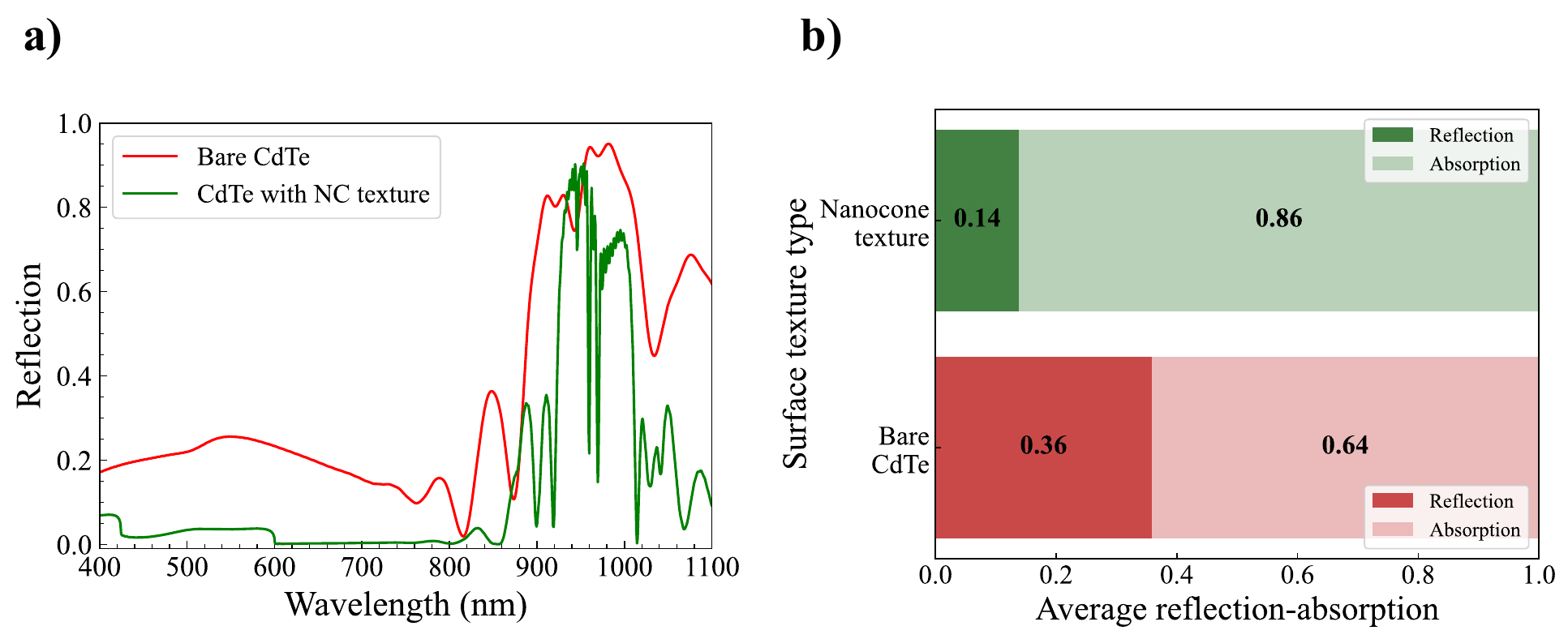}
 \caption{(a) Reflection spectra, (b) average reflection-absorption for bare CdTe TFSCs and NC
textured CdTe over the wavelength range of 400 nm to 1100 nm.}
 \label{reflectance}
\end{figure*}

V\textsubscript{oc} is the output voltage of a solar cell under open-circuit condition, i.e., when the current through the solar cell is 0 A. Thus, it is also the maximum cell voltage when no photocurrent is generated due to the dark current and photocurrent generated being equal. It is calculated using the formula \cite{num_analysis},
\begin{equation}
\label{voc}
V\textsubscript{oc}=\frac{nkT}{q}ln\left(\frac{I\textsubscript{L}}{I\textsubscript{0}}+1\right)
\end{equation}
where \textit{n} is the ideality factor, \textit{k} is the Boltzmann constant, \textit{T} is the temperature in Kelvin, I\textsubscript{L} is the light-generated current, and I\textsubscript{0} is the dark saturation current.

Fill factor is ratio of the maximum possible power output of a solar cell to the product of its respective J\textsubscript{sc} and  V\textsubscript{oc}. It is also an indication of the quality of the solar cell, since a larger fill factor indicates a more optimal performance of the solar cell. It is calculated using the formula \cite{num_analysis}, 
\begin{equation}
\label{fill_factor}
FF=\frac{P\textsubscript{max}}{J\textsubscript{sc}\text{*}V\textsubscript{oc}}
\end{equation}
PCE is the key parameter determining how much of the incident power on the solar cell is converted to useful electrical energy, and it is dependent on the other important electrical parameters – J\textsubscript{sc}, V\textsubscript{oc}, and fill factor – as shown below \cite{num_analysis},
\begin{equation}
\label{efficiency}
PCE=\frac{FF\text{*}J\textsubscript{sc}\text{*}V\textsubscript{oc}}{P\textsubscript{in}}
\end{equation}
where P\textsubscript{in} is the total power incident on the solar cell, typically 100 mW/cm\textsuperscript{2} according to the AM1.5G solar model \cite{AM1.5}.

\section{Results and discussions}

This section systematically optimizes the morphology of the CdS-CdTe NC texture and embedded spherical Ge NP followed by their underlying enhancement mechanism. It has been shown through grating analysis and Mie analysis that texture and NP are desirable options for achieving broadband light absorption. This section extends further to compare the opto-electronic performance parameters of bare CdTe TFSCs, NC textured CdTe and NC textured CdTe with embedded Ge NPs, respectively. The final subsection includes polarization and angle of incidence analyses of the optimal structure.

\subsection{Nanocone texture optimization}

\begin{figure*}[!t]
 \centering
 \includegraphics[width=\textwidth]{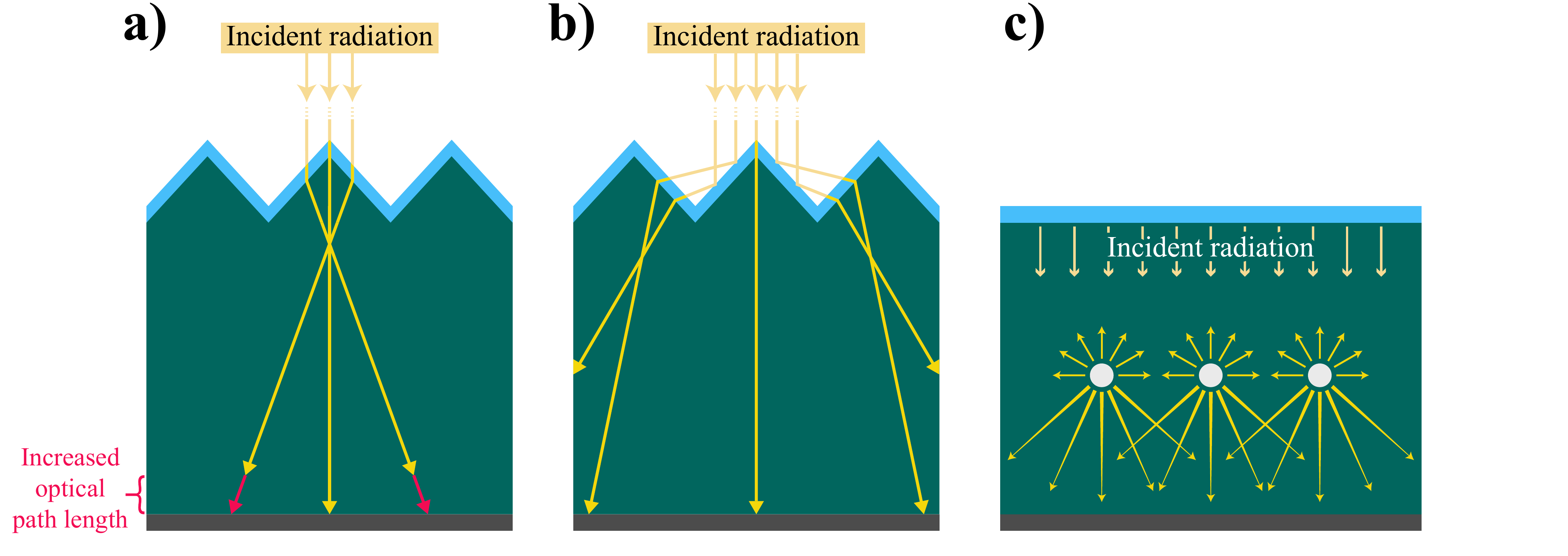}
 \caption{Depiction of (a) path length increase in the presence of NC texture, (b) reflection of light at the top surface and subsequent transmission of the reflected light into the CdTe absorber layer in the presence of NC texture, (c) high-intensity light scattering of the embedded Ge NP at the CdTe absorber layer.}
 \label{scatter}
\end{figure*}

The initial target of this study is to achieve an optimal texture configuration for which the reflectance is minimal, and the absorption is maximum for the NC textured CdTe TFSCs. To achieve this, NC shaped texture of both CdS-CdTe layers has been optimized by varying the cone height and diameter at the same time. Electrical performance parameters like J\textsubscript{sc}, V\textsubscript{oc}, fill factor, and PCE are all taken into consideration to achieve the optimal morphology for the NC texture.

Fig. \ref{texture_optimization} shows the performance variation of J\textsubscript{sc}, V\textsubscript{oc}, fill factor, and PCE when the NC height and diameter are varied. The cone height was varied from 50 nm to 400 nm with an increment of 50 nm whereas the base diameter was varied from 300 nm to 900 nm with a 100 nm increment. It is worth mentioning that both parameters were varied at the same time to find the optimal point in the two-dimensional solution space. By changing the structural parameters of NC texture, J\textsubscript{sc} can be significantly improved compared to bare CdTe TFSCs \cite{MicNanoLig}. From Fig. \ref{texture_optimization}(a), one can easily infer that, for a fixed base diameter, J\textsubscript{sc} tends to increase with the increase of cone height. An increase in height also means an increase in aspect ratio which can be defined as a ratio of height to base diameter. The highest J\textsubscript{sc} of 32.09 mA/cm\textsuperscript{2} (see Table S2) was achieved for a cone base diameter and height of 600 nm and 400 nm, respectively, making the aspect ratio equal to 0.67. Such aspect ratio is found to be a factor in achieving the maximum J\textsubscript{sc} for other solar cells with similar types of top surface texturing \cite{AbsEnhGaAs}. Compared to bare CdTe TFSCs, this is a 31.93\% enhancement (J\textsubscript{sc} for bare CdTe is 24.33 mA/cm\textsuperscript{2}). The reason for this increase can be attributed to the increase in diffused transmission of light. That is to say, with the increase in height, surface texture contributes to the increase of light transmission and reduces reflective losses \cite{ReaGlaSur}. Unlike J\textsubscript{sc}, the V\textsubscript{oc} change in Fig. \ref{texture_optimization}(b) is somewhat different. The highest achievable V\textsubscript{oc} was 930.4 mV for an NC base diameter and height of 500 nm and 50 nm. Compared to bare CdTe TFSCs (V\textsubscript{oc} = 978 mV), this is only 5.04\% lower.  Clearly, surface recombination plays a crucial role when determining the V\textsubscript{oc} since any textured surface will eventually increase the surface coverage and so can also be expected to increase surface recombination rates. Fig. \ref{texture_optimization}(c) shows the maximum fill factor of 0.8355 was achieved for the highest aspect ratio point where the base diameter of the cone is minimum (300 nm) and height is maximum (400 nm). Adding texture has significantly improved the quality of the solar cell with respect to base CdTe TFSCs (FF for bare is 0.6455) and can be credited to the improved carrier collection. Finally, the PCE of the solar cell with the variation of NC height and base diameter is found to follow the same trend as J\textsubscript{sc} because the change in J\textsubscript{sc} is significant and contributes more to determining the PCE compared to the other two parameters i.e. V\textsubscript{oc} and fill factor. The highest PCE of 24.62\% was achieved for NC base diameter and height of 600 nm and 400 nm respectively, an 60.28\% increase compared to the baseline structure (see Table S2).

\subsubsection{Reflectance spectra analysis for nanocone texture}

\begin{figure*}[!t]
 \centering
 \includegraphics[width=\textwidth]{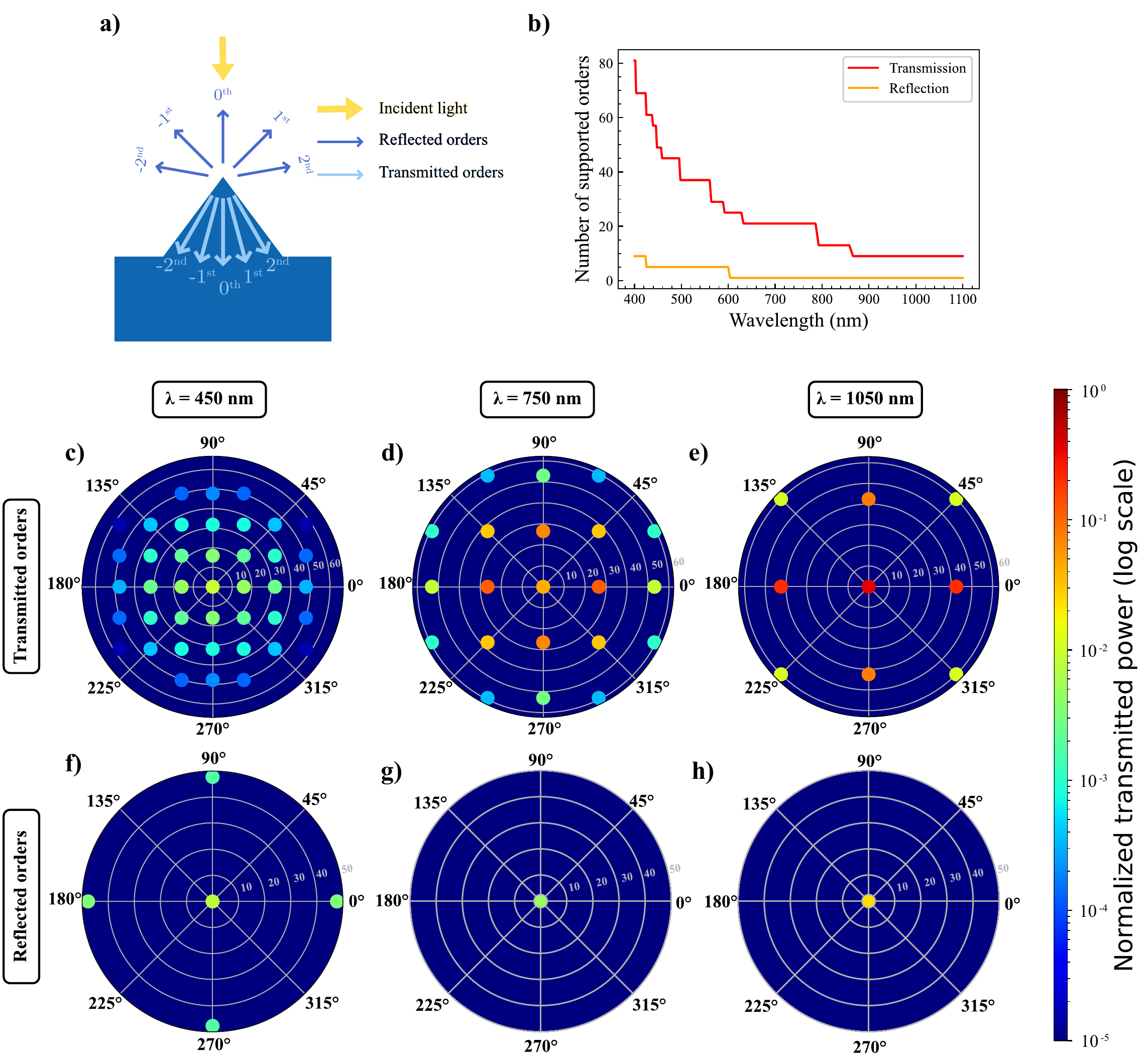}
 \caption{(a) Formation of transmitted and reflected orders from NC texture, (b) number of supported transmitted and reflected orders; propagation direction and strength of transmitted orders at (c) $\lambda$ = 450 nm, (d) $\lambda$ = 750 nm, (e) $\lambda$ = 1050 nm; reflected orders at (f) $\lambda$ = 450 nm, (g) $\lambda$ = 750 nm, (h) $\lambda$ = 1050 nm.}
 \label{diffration_orders}
\end{figure*}

In the previous subsection, 600 nm and 400 nm were found to be the respective optimal base diameter and height of NC texture to achieve the highest PCE as well as J\textsubscript{sc}. In this subsection, reflectance spectra will be analyzed to understand the performance improvement. 

Fig. \ref{reflectance}(a) depicts the reflectance spectra for bare CdTe TFSCs and CdTe TFSCs modified with the optimal NC structure obtained in the previous section. Reflection from the top surface of the solar cell mainly occurs due to the refractive index mismatch of two different layers \cite{AbsEnhUltra}. It can be realized from Fig. \ref{reflectance}(a) that optimal NC texture can suppress reflection from the top surface over almost all the wavelengths from 400 nm to 1100 nm. This is due to the fact that an optimally textured top surface works as an effective medium of graded refractive index as the cross-section starts from zero (at the apex of NC) and reaches the maximum (at the base of NC) as the light goes to the substrate. This allows index matching between glass and CdS-CdTe layers and thus reduces reflection from the top surface \cite{AbsEnhGaAs}. Additionally, NC texture breaks the incoming wavefront’s uniformity and focuses it into a specific direction to be absorbed \cite{UsiPlasNano}. Out of all the wavelengths, between 400 nm to 800 nm (absorption window of CdTe), the average reflection came down to almost zero, owing to the pitch or base diameter of the optimal NC. As discussed in the previous theoretical studies, in terms of the light trapping prospect of the texture structure, the base diameter or pitch has to be close to the targeted wavelength \cite{FunLimLig, FunLimNano}. Since surface texture is suitable for applications in the visible wavelength range, a 600 nm base diameter of NC is optimal and aligns with the target wavelength (midpoint of visible wavelengths). Fig. \ref{reflectance}(b) portrays the side-by-side average reflection and absorption for the Bare CdTe TFSCs and CdTe TFSCs modified with optimal NC texture. Flat-surfaced bare CdTe TFSCs absorb almost 64\% of incoming photons, whereas the remaining 36\% are lost as reflective losses. Optimal NC texture on the other hand brings down the reflective losses to only 14\%, allowing 86\% of the photons to be absorbed by the absorber layer. Optimal NC texture scatters incoming transmitted light to an oblique angle causing them to travel longer than it would have if the surface were flat \cite{NanoRearAg} as visually illustrated in Fig. \ref{scatter}(a). Even if the incident light gets reflected by one cone surface, it might reflect the light at the second cone surface for it to be transmitted into the absorber layer, resulting in more absorption as shown in Fig. \ref{scatter}(b). Moreover, this scattered light then again might be scattered back into the absorber layer by the back reflector and front texture. This multiple internal reflection largely increases the optical path length and so the absorption \cite{AbsEnhGaAs}. In addition to texture, NP later contributes to scattering light in the forward direction as illustrated in Fig. \ref{scatter}(c) to aid unabsorbed photons in being absorbed. Further analysis in support of these mechanisms is discussed in the later sections.

\subsubsection{Nanocone texture grating analysis}

\begin{figure*}[!t]
 \centering
 \includegraphics[width=\textwidth]{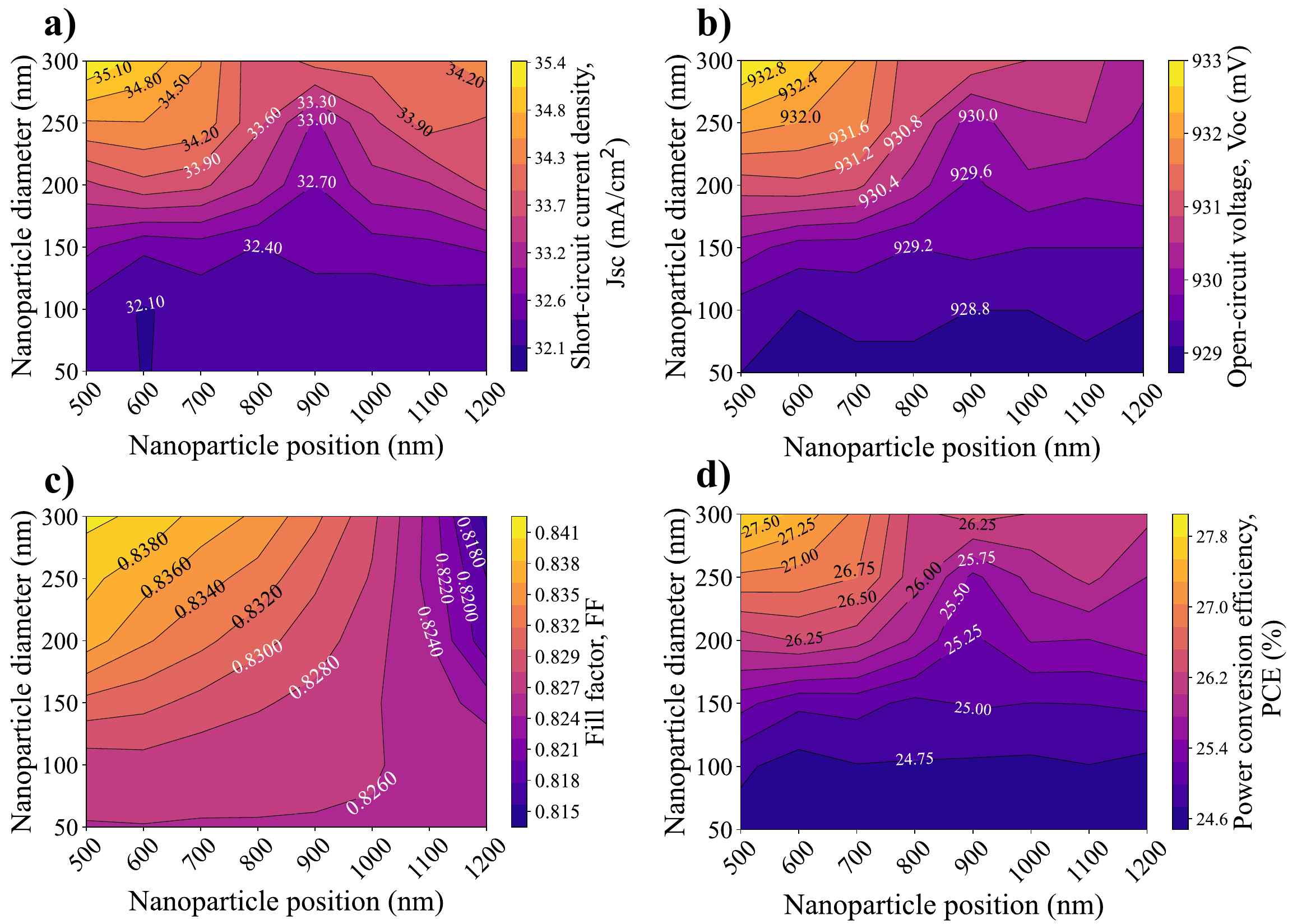}
 \caption{Contour plot of photovoltaic parameters i.e. (a) short-circuit current density (J\textsubscript{sc}), (b)
open-circuit voltage (V\textsubscript{oc}), (c) fill factor (FF), and (d) power conversion efficiency (PCE) as a
function of Ge NP position (z\textsubscript{Ge}) and diameter (d\textsubscript{Ge}). J\textsubscript{sc}, V\textsubscript{oc}, fill factor, and PCE for bare CdTe TFSCs are 24.33 mA/cm\textsuperscript{2}, 978 mV, 0.6455, and 15.36\%, respectively.}
 \label{nanoparticle_optimization}
\end{figure*}

Nanophotonic texture whose dimension is within the order of the wavelength can not only reduce reflection from the top surface but also trap light in the absorber layer through diffraction by working as a diffraction grating \cite{NanoLigTrap, WavSelAbs}. NC texture designed in this study also falls within the category of two-dimensional diffraction grating, and so a grating analysis is essential to understand the mechanism of absorption enhancement.  Any periodically arranged uniform texture can transmit and reflect the diffracted light to create constructive interference of light. These constructive interferences of light are called transmitted diffraction orders and reflected diffraction orders as shown in Fig. \ref{diffration_orders}(a). For two-dimensional grating, the diffraction equation can be written as \cite{InvDesPol},
\begin{equation}
\sin\theta_{m,n} \cos\varphi_{m,n} = \sin\theta \cos\varphi + \frac{m \lambda}{D_x}
\end{equation}
\begin{equation}
\sin\theta_{m,n} \sin\varphi_{m,n} = \sin\theta \sin\varphi + \frac{n \lambda}{D_y}
\end{equation}
where $\theta$ and $\varphi$ denote the polar and azimuth angle of the incident light, respectively, $\lambda$ is the wavelength for which diffraction order location will be calculated, $\theta$\textsubscript{m,n} and $\varphi$\textsubscript{m,n} are respective polar and azimuth angular location of (m,n) diffraction order, and finally D\textsubscript{x} and D\textsubscript{y} are the grating period in the x and y-axis directions. For a normal incident light and an equal grating period in both axis directions, the above two equations can be reduced to the following,
\begin{equation}
\sqrt{m^2 + n^2} = \frac{D \sin \theta_{m,n}}{\lambda}
\end{equation}
Far-field effect originating from a diffraction grating can enhance light absorption by increasing the path length and focusing light into specific diffraction modes \cite{UsiPlasNano}. To visually understand each supported diffraction order, a far-field semi-sphere is often used to locate all the orders along with their angular location and strength. Fig. \ref{diffration_orders}(c-h) depicts the polar and azimuth angular location of all supported orders for the optimal NC texture grating with a color bar to indicate their strength. To be specific, while Fig. \ref{diffration_orders}(c), (d), and (e) shows all the transmitted orders at wavelengths of 450, 750, and 1050 nm, respectively, Fig. \ref{diffration_orders}(f), (g), and (h) shows all the reflected orders for the same wavelengths. As seen in Fig. \ref{diffration_orders}(b), the number of supported transmitted orders is significantly higher than the supported reflected orders, indicating the optimal NC texture can reduce the reflective losses and simultaneously contribute to coupling light into the absorber layer through the far-field effect. The number of supported orders seems to be going down with the increase in wavelength meaning the NC texture has more desirable properties in the shorter wavelength region compared to longer wavelengths. Also, the transmitted orders are found to be focused at the center of the hemisphere at 450 nm wavelength and expanding at later two wavelengths meaning the path length increases at longer wavelengths and thus improves photon capture probability.

\subsection{Embedded Ge nanoparticle optimization}

In this part, the focus is to optimize the spherical Ge NP size, i.e., diameter and its position. It is important to highlight that, the NP optimization was done on the previous optimized texture structure where both the CdS-CdTe surface layers were textured as NCs with a base diameter and height of 600 nm and 400 nm, respectively. Since the Ge NP has been embedded into the CdTe absorber layer, its position is defined as the depth (Z\textsubscript{Ge}) from the top CdS flat surface. Both the Ge NP’s diameter and depth have been varied at the same time and J\textsubscript{sc}, V\textsubscript{oc}, fill factor, and PCE are all taken into consideration to achieve the optimal position and size for Ge NP. 

Fig. \ref{nanoparticle_optimization} illustrates the performance variation of J\textsubscript{sc}, V\textsubscript{oc}, fill factor, and PCE when the Ge NP diameter and position are varied. The NP diameter was varied from 50 nm to 300 nm with an increment of 50 nm whereas the position was varied from 500 nm to 1200 nm from the top of the CdS surface in 100 nm increments. It is noteworthy to mention that the increase in position means the NP is embedded deeper into the CdTe absorber layer. Here, Fig. \ref{nanoparticle_optimization}(a), (b), and (d) reveal a similar pattern where better J\textsubscript{sc}, V\textsubscript{oc}, and PCE results were achieved when the Ge NP size was bigger compared to the smaller ones. In those cases, the NP position matters less than the NP size. The reason can be attributed to the fact that the Ge nanosphere shows size-dependent light scattering properties spanning the whole visible to NIR region and the scattering intensity increases with the increase in size \cite{DirScaGer}. However, in the case of the fill factor, the trend is slightly different, and one may conclude that the NP position is prominent in terms of change compared to the NP size. Yet all the performance parameters including the fill factor demonstrate the highest value for the NP’s diameter and depth of 300 nm and 500 nm, respectively. The highest possible J\textsubscript{sc}, V\textsubscript{oc}, fill factor, and PCE was found to be 35.38 mA/cm\textsuperscript{2}, 933.2 mV, 0.8406, and 27.76\%, respectively (see Table S3). When compared with the only optimal NC textured structure, this optimized Ge NP with NC texture adds more than 20\% PCE enhancement while the performance of the rest of the parameters was also elevated as shown in Table \ref{performance}. This supports the fact that composite light trapping can provide a means to achieve highly efficient CdTe TFSCs.

\subsubsection{Ge nanoparticle Mie scattering analysis}

Mie theory can effectively use Maxwell’s equations to calculate the scattering of a spherical particle illuminated with a monochromatic plane wave in a homogenous medium \cite{MicNanoLig}. Typically, illuminated NP within the range of incident light wavelength typically shows strong forward light scattering. Such strong scattered light can be harnessed by photosensitive material like solar cells and increase its absorption. Furthermore, Mie's theory can also provide the angular distribution of scattered light that varies with the change in NP size, refractive index, and incident light wavelength \cite{ExpTecAgg}. Two of the main parameters that are calculated using Mie theory are scattering and absorption cross-section often denoted as C\textsubscript{scatt} and C\textsubscript{abs}, respectively \cite{LigScaCal}. The scattering cross-section is defined as the total scattered power by the spherical particle with respect to the power per unit area of the incident light \cite{LigScaCal, LigManPho}. Similarly, the absorption cross-section is defined as the same except this time total absorbed power by the spherical particle takes the numerator position of the ratio. Often times these two parameters are normalized to the geometrical cross-section of the scattering or absorbing objects \cite{LigScaCal}. The scattering and absorption cross-sections of a spherical NP can be calculated using the following formulas \cite{PlasEnhPho},
\begin{equation} C_{\textsubscript{scatt}} = \frac{1}{6\pi} k\textsuperscript{4} \left| \alpha_{\textsubscript{sph.}} \right|\textsuperscript{2}
\end{equation}
\begin{equation}
C_{\textsubscript{abs}} = k \, \text{Im}[\alpha_{\textsubscript{sph.}}] \end{equation}
where k(=2$\pi/\lambda$) represents the wavenumber of the incident light and $\alpha$\textsubscript{sph.} is the polarizability of the spherical particle expressed as,
\begin{equation}
\alpha_{\text{sph.}} = 3 V \left[ \frac{\varepsilon_{\text{np}} - \varepsilon_{\text{sm}}}{\varepsilon_{\text{np}} + 2\varepsilon_{\text{sm}}} \right]
\end{equation}
Here, V symbolizes the volume of the particle, and $\varepsilon$\textsubscript{np}, and $\varepsilon$\textsubscript{sm} are the permittivity of the NP and surrounding medium, respectively. 

To normalize against the geometrical cross sections, the following formulas are used \cite{EnhOptAbs}, 
\begin{equation} 
Q_{\text{abs}} = \frac{C_{\text{abs}}}{\pi r^2}
\end{equation}
\begin{equation}
Q_{\text{scatt}} = \frac{C_{\text{scatt}}}{\pi r^2}
\end{equation}
Here, r is the radius of the spherical particles. 

Fig. S2(a-b) present the normalized scattering and absorption cross sections, respectively, for Ge NPs with diameters between 50 nm and 300 nm, with an increment of 50 nm. It is evident from these two figures that except for 50 nm diameter Ge NP, normalized scattering cross sections are higher compared to the absorption cross sections. To effectively measure this parameter, the scattering to absorption cross-section ratio was plotted in Fig. S2(c). It is clear that as the size increases, Ge NPs tend to scatter more light than it absorbs which is consistent with the previously reported pieces of literature \cite{PlasEnhPho, ResOptAbs, DirScaGer}. The highest absorption cross-section ratio was achieved for the 300 nm diameter Ge NP, which is close to 4 at above around $\lambda$ = 1000 nm. Furthermore, in contrast to smaller Ge NP, as the diameter of the Ge NP increases, red shifting is observed in the scattering cross sections. Finally, Fig. S2(d) depicts the absorption, scattering, and extinction cross sections (sum of the absorption and scattering cross sections) for 300 nm diameter Ge NP.

In the context of this work, these results justify why 300 nm sized Ge NP and 500 nm NP position resulted in the highest J\textsubscript{sc}, V\textsubscript{oc}, fill factor and PCE. As 300 nm Ge NP scatters more light compared to the other NP sizes, it increases the possibility of low-frequency photons being absorbed by the CdTe absorber layer since most of the photons of NIR range light remain unabsorbed \cite{EnhOptAbs}. This might be due to the fact that embedded NP in the absorber layer redistributes the optical field to result in a homogenization of the optical field (HOF) \cite{HomOptFie}. This can significantly increase the carrier generation at longer wavelengths. Furthermore, the placement of the NPs is immensely important considering their light-scattering properties and the direction in which they scatter light. For example, if the NP is scattering light in the downward direction and placed at the bottom of the absorber layer, it would not effectively contribute to absorption enhancement, since scattered light does not have enough absorber material to be coupled with. Thus, when the optimal Ge NP is placed at 500 nm depth i.e. closer to the surface, its forward scattered light can pass through a sufficiently large portion of CdTe absorber material, giving a chance to generate more charge carriers. 

\begin{figure*}[!t]
 \centering
 \includegraphics[width=\textwidth]{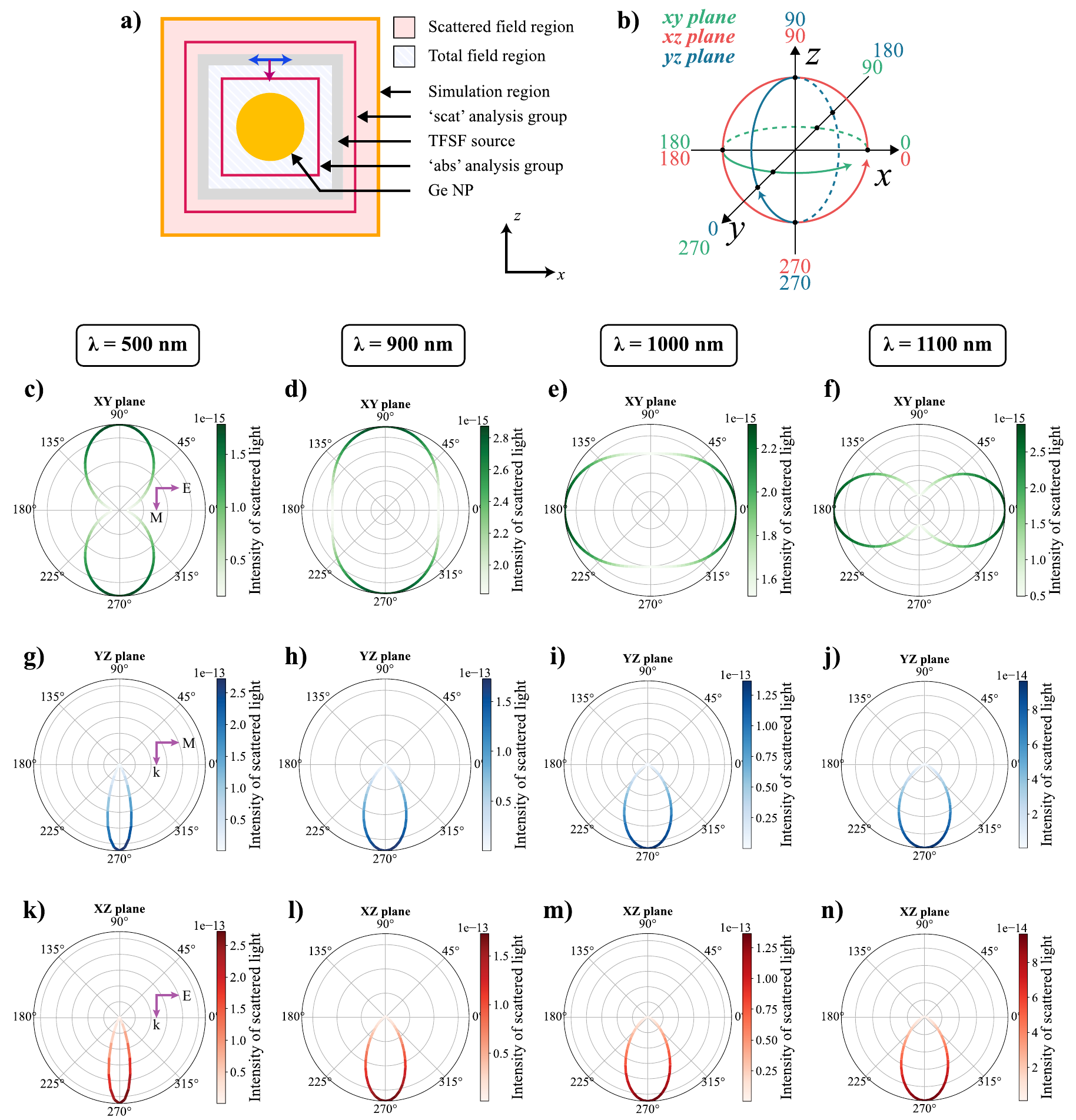}
 \caption{(a) Simulation setup for Mie scattering analysis of 300 nm diameter Ge NP with TFSF source. (b) Polar angle definition in each of the x-y, y-z, and x-z planes with respect to the x, y, and z axes; Polar plots showing the far-field scattered light in the x-y (c-f), y-z (g-j) and x-z (k-n) planes for the wavelengths of $\lambda$ = 500 nm, 900 nm, 1000 nm, and 1100 nm as denoted at the top of each figure column grid. Here, 'k', 'E', and 'M' in (c), (g), and (k) denote the direction of the incident light, electric field, and magnetic field polarization respectively.}
 \label{polar_plot}
\end{figure*}

\begin{table}[!b]
\small
  \caption{\ Performance Comparison of bare CdTe TFSCs, NC textured CdTe and NC textured CdTe with embedded Ge NP, respectively}
  \label{performance}
  \begin{tabular*}{0.48\textwidth}{@{\extracolsep{\fill}}h{0.17\textwidth}h{0.07\textwidth}h{0.05\textwidth}h{0.05\textwidth}h{0.05\textwidth}}
    \hline
         \textbf{Configuration}&  \textbf{J\textsubscript{sc} (mA/cm\textsuperscript{2})}& \textbf{V\textsubscript{oc} (mV)}& \textbf{Fill Factor}& \textbf{PCE (\%)}\\
    \hline
         Bare CdTe TFSCs& 24.33& 978& 0.6455& 15.36\\
         NC textured CdTe& 32.09& 928.7& 0.8259& 24.62\\
         NC textured CdTe with embedded Ge NP& 35.38& 933.2& 0.8406 &27.76\\ 
    \hline
  \end{tabular*}
\end{table}

\begin{figure*}[!t]
 \centering
 \includegraphics[width=\textwidth]{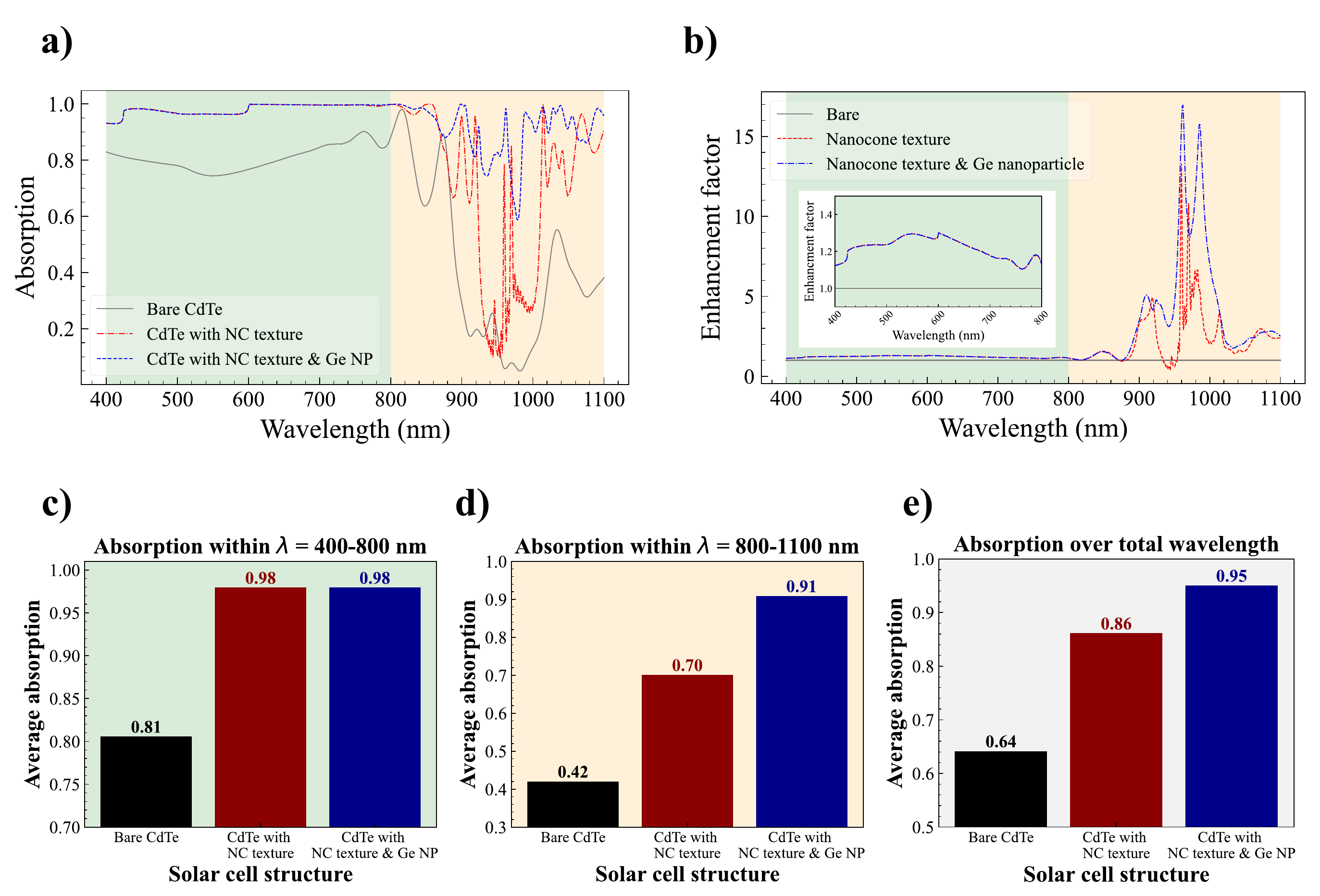}
 \caption{(a) Absorption spectra, (b) enhancement factor, (c) average absorption within 400-800
nm wavelength range, (d) average absorption within 800-1100 nm wavelength range, (e) average absorption over the total wavelength (400-1100 nm) range for bare CdTe TFSCs, NC textured CdTe and NC textured CdTe with Ge NP.}
 \label{absorption}
\end{figure*}

Apart from quantitively measuring the scattered light through scattering and absorption cross-sections, a far-field scattering plot is also important considering it can provide fundamental insight into the direction of the scattered light. Ge being a high-indexed semiconductor NP can trap light via far-field light coupling of scattered light and nearfield coupling of localized light \cite{EnhPhoPer}. To calculate far-field scattering, 300 nm diameter Ge NP was illuminated with a TFSF source as shown in Fig. \ref{polar_plot}(a) and the direction of incident light is in the negative z direction. Fig. \ref{polar_plot}(b) defines the polar angle in each plane with respect to the x, y, and z axes. Fig. \ref{polar_plot}(c-f) presents far-field scattered light in the x-y plane, Fig. \ref{polar_plot}(g-j) for the y-z plane, and Fig. \ref{polar_plot}(k-n) for the x-z plane along with a color bar representing the intensity of light. All these far-field angular distributions are plotted against four different wavelengths of 500, 900, 1000, and 1100 nm marked at the top of each column in Fig. \ref{polar_plot} grid. The far-field radiation pattern at a wavelength of 500 nm provides the baseline pattern, where the contribution of Ge NPs is minimal compared to the near-infrared (NIR) wavelengths of 900, 1000, and 1100 nm, where the Ge NPs significantly contribute to light absorption. From Fig. \ref{polar_plot}(g-n), one can deduce that the angular far-field distribution of the y-z and the x-z plane shows the scattered light is only focused in the z-direction (at 270$^{\circ}$) meaning almost all the light is scattered in the forward direction and and there is very little (if any) backward reflection of the incident light from the surface of the CdTe TFSCs. On the other hand, Fig. \ref{polar_plot}(c-f) shows an alternating dipole resonance between the direction of electric and magnetic fields with the change in wavelengths for the x-y plane. Subwavelength Ge NP can demonstrate both electric and magnetic dipole resonance as described in previous literature \cite{GeDipole}  The direction of both electric and magnetic dipole lobes is transverse to the direction of incident light. However, the light intensity of this dipole response (10\textsuperscript{-15} order) is comparatively less than the forward scattered light (10\textsuperscript{-13} order) to the direction of incident light in the y-z and x-z plane. In the y-z and the x-z plane, it can be seen that with the increase of incident light wavelength, the light scatters forward (at 270$^{\circ}$) at a wider angle. This eventually increases the optical path length with the help of the metal back reflector leading to enhanced absorption \cite{UsiPlasNano}. This further validates why the absorption of CdTe TFSCs is significantly enhanced in the wavelength range of 800 nm to 1100 nm as shown in the absorption plots in the later sub-sections.

\begin{figure*}[!h]
 \centering
 \includegraphics[height=23cm]{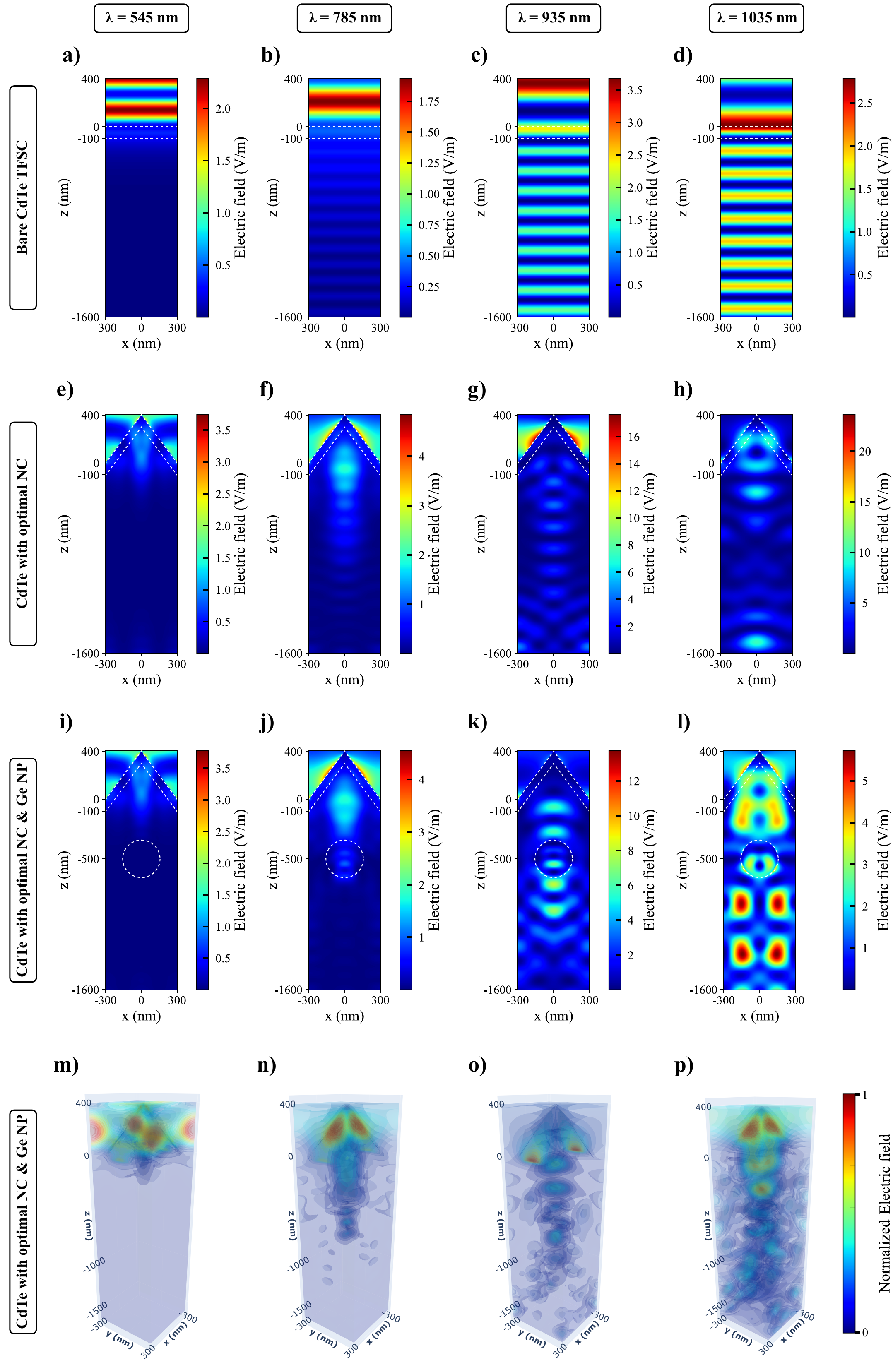}
 \caption{Electric field distribution plot for bare CdTe TFSCs (a-d), NC textured CdTe (e-h), and
NC textured CdTe with Ge NP (i-l) for the wavelength $\lambda$ = 545 nm, 785 nm, 935 nm, and 1035 nm. (m-p) Normalized 3D electric field distribution plot for NC textured CdTe with Ge NP.}
 \label{electric_field}
\end{figure*}

\subsection{Comparative analysis of bare CdTe TFSCs, nanocone textured CdTe, and nanocone textured CdTe with Ge NP}
\subsubsection{Broadband absorption spectra analysis}

\begin{figure*}[!t]
 \centering
 \includegraphics[width=\textwidth]{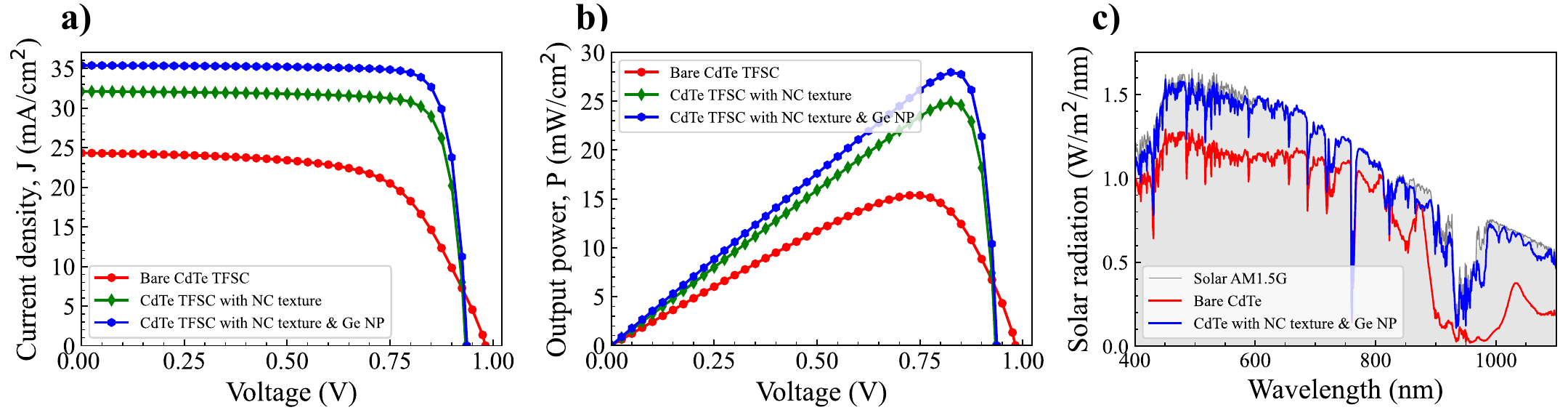}
 \caption{(a) J-V curve, (b) P-V curve for bare CdTe TFSCs, NC textured CdTe and NC textured CdTe with Ge NP, (c) spectral intensity received and absorbed for bare CdTe TFSCs and proposed CdTe TFSC modified with NC structure and embedded with Ge NP with respect to AM 1.5G incident solar radiation.}
 \label{PV_IV}
\end{figure*}

To understand the individual and combined performance enhancement by the optimal NC texture and Ge NP, absorption spectra over the whole wavelength region of 400 nm-1100 nm have been shown in Fig. \ref{absorption}(a). For better understanding, the whole wavelength range has been divided into two regions. The 400 nm to 800 nm wavelength region and 800 nm to 1100 nm wavelength region are shaded with two different colors, considering short and long wavelength regions, respectively in terms of solar cell absorption. In the short wavelength region (400 nm - 800 nm), the average absorption of the CdTe absorber layer without any light trapping is already over 80\% (Fig. \ref{absorption}(c)) due to its high absorption coefficient. Yet the absorption of this region is even further elevated with the help of the NC surface texture and reaches 98\% as seen in Fig. \ref{absorption}(c). The optimal NC surface texture (d\textsubscript{cone} = 600 nm, h\textsubscript{cone} = 400 nm) not only reduces the light reflection but also work as a medium of light trapping to redirect light into the absorber layer at this wavelength range. Since most of the light is absorbed within a few hundred nanometers of CdTe layers, Ge NP plays little to no role in increasing absorption at a short wavelength range. However, when it comes to the long wavelength region (800 nm - 1100 nm), 300 nm diameter Ge NP placed 500 nm below the CdS top surface along with optimal NC texture can raise the average absorption to 91\%, which is more than double the average absorption of Bare CdTe TFSCs (42\%) as shown in Fig. \ref{absorption}(d). Moreover, the contribution of the optimal NC texture can not be denied in the long wavelength range which can alone increase average absorption to 70\% without Ge NPs. This implies that the optimal NC surface texture can work as an excellent anti-reflection medium over all the wavelength ranges along with providing light trapping at the same time. Considering the total wavelength of 400 nm to 1100 nm, CdTe TFSCs with optimal NC surface texture and embedded Ge NP can provide a 95\% average absorption, suggesting the broadband absorption compared to bare CdTe (64\%) or with only the NC texture (86\%) as seen in Fig. \ref{absorption}(e). To compare how much the absorption has been elevated compared to bare CdTe TFSCs, the enhancement factor h($\lambda$) corresponding to each wavelength have been plotted in Fig. \ref{absorption}(b). The following equation is used to calculate the enhancement factor,
\begin{equation}
\small
h(\lambda) = \frac{\text{absorption of CdTe TFSCs with texture and/or NP at } \lambda}{\text{absorption of planar CdTe TFSCs at } \lambda}
\normalsize
\end{equation}
Fig. \ref{electric_field}(b) reveals that the highest enhancement has taken place in the long wavelength range due to the weak absorption of CdTe TFSCs at this range. Compared to only NC texture configuration, CdTe with optimal NC texture with Ge NP can provide higher enhancement in this wavelength region. The inset figure in Fig. \ref{absorption}(b) shows the enhancement factor in the short wavelength region. The reason the enhancement factor tends to remain between 1.1 to 1.4 is that the absorption of bare CdTe TFSCs is already high in this region. Similar to average absorption, J\textsubscript{sc} contribution over certain wavelength ranges is plotted in Fig. S3(a-c). For $\lambda$ = 400-800 nm, J\textsubscript{sc} elevates from 18.77 mA/cm\textsuperscript{2} to 22.78 mA/cm\textsuperscript{2} in a similar manner of average absorption for NC texture with and without Ge NP as seen in Fig. S3(a). However, NC texture increases J\textsubscript{sc} to almost double (6.63 mA/cm\textsuperscript{2}) and embedded Ge NP with NC texture to more than three times (11.65 mA/cm\textsuperscript{2}) compared to bare CdTe TFSCs (3.45 mA/cm\textsuperscript{2}) when $\lambda$ = 800-1100 nm, as shown in Fig. S3(b).

\begin{figure*}[!t]
 \centering
 \includegraphics[height=6cm]{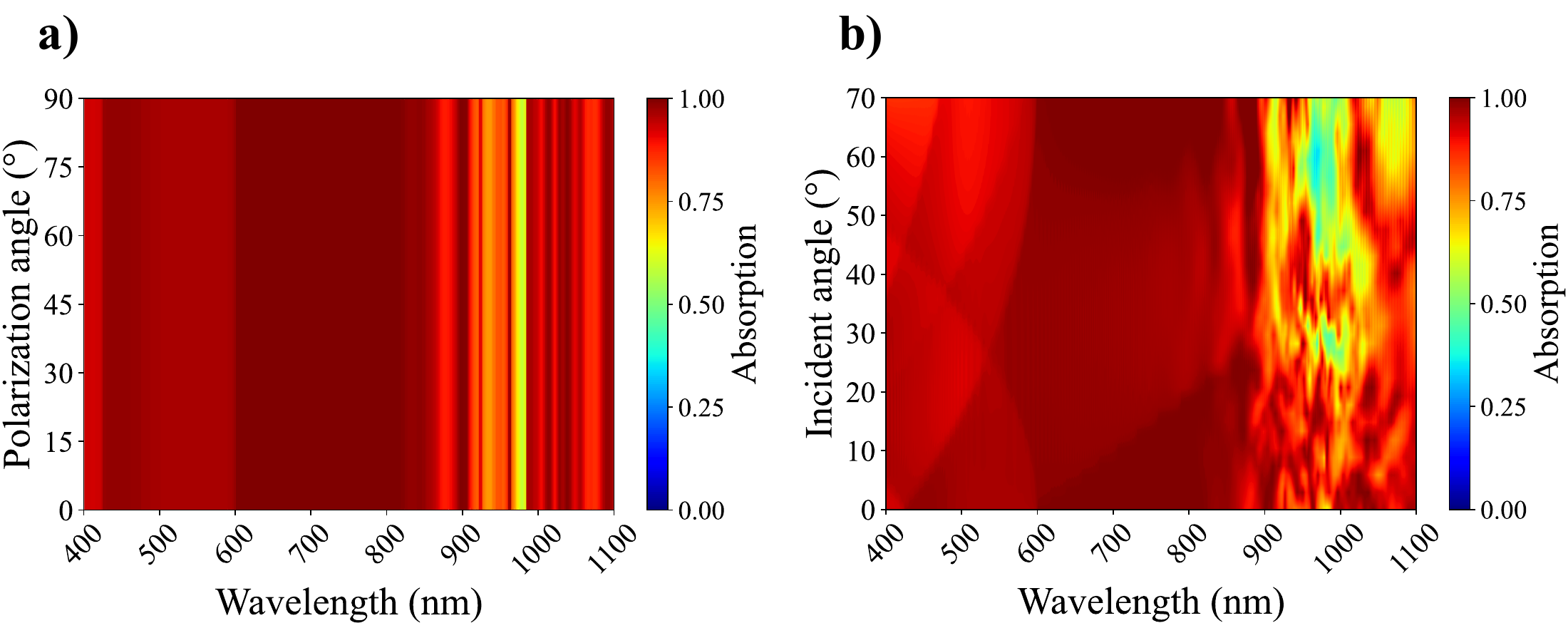}
 \caption{Absorption as a function of (a) polarization angle, (b) incident angle for optimal NC textured CdTe with embedded Ge NP.}
 \label{polarization_incident_angle}
\end{figure*}

\begin{table*}[!t]
\small
  \caption{\ Performance comparison of the proposed structure with recently studied CdTe TFSCs}
  \label{comparison}
  \begin{tabular*}{\textwidth}{@{\extracolsep{\fill}}h{0.33\textwidth}h{0.1\textwidth}h{0.08\textwidth}h{0.05\textwidth}h{0.05\textwidth}h{0.04\textwidth}h{0.07\textwidth}}
    \hline
         \textbf{Solar cell structure}&  \textbf{Light trapping technique}& \textbf{J\textsubscript{sc} (mA/cm\textsuperscript{2})}& \textbf{V\textsubscript{oc} (V)}& \textbf{Fill Factor}& \textbf{PCE (\%)}& \textbf{Reference}\\
    \hline

         SiO\textsubscript{2}/Si(DBR)/ZnTe/CdTe/CdS/ZnO/FTO/glass&  SiO\textsubscript{2}/Si (DBR)& 10.98& 1.201& 0.82& 10.39& \cite{2020SiSiO2DBR}\\        FTO/SnO\textsubscript{2}/CdS/CdTe/MoO\textsubscript{3}/(MgF\textsubscript{2}/MoO\textsubscript{3})\textsuperscript{4}&  1D-PC& 17.23& 0.960& 0.63& 10.47& \cite{2022_1DPC}\\
         ITO/CdS/CdTe/Ag& Micro-texturing& 24.29& 0.798& 0.57& 10.98& \cite{2024MicroTexture}\\
         Au/ZnO/CdS/CdTe/Au& ZnO Nanopillar& 16.90& 1.008& 0.75& 12.60& \cite{2021NanoPillar}\\
         Ag/CdS/CdTe/Al(coating)/glass(grating)/Ag& Nano-grating& 30.52& 1.020& 0.75& 23.48& \cite{2024NG}\\
         ZnS/CdTe/BSF/DBR& Si/Al\textsubscript{2}O\textsubscript{3} DBR& 25.04& 1.065& 0.88& 23.94& \cite{2019DBR}\\         
         ITO/CdS/CdTe/Au/Ni& NW SC \& ITO FS& 41.56& 0.871& 0.73& 26.45& \cite{2022FS}\\
         Glass/Ag/CdS/CdTe/Ge/Ag&  NC texture \& Ge NP& 35.38& 0.933& 0.84& 27.76& This work\\
    \hline
  \end{tabular*}
\end{table*}

\subsubsection{Electric field analysis for TM mode polarization}

Fig. \ref{electric_field}(a-p) demonstrates the electric field distribution plot for the wavelengths of 545 nm, 785 nm, 935 nm, and 1035 nm for all three configurations. Since electric field distribution for all the wavelengths can not be practically shown in this work, these four wavelengths are chosen as they provide more light-matter interaction than other wavelengths do. The four columns of the Fig. \ref{electric_field} grid correspond to the four wavelengths of 545 nm, 785 nm, 935 nm, and 1035 nm, respectively. The first three rows belong to the bare CdTe TFSCs, with optimal NC-textured CdTe, and finally, NC-textured CdTe with embedded Ge NP, respectively. Since it is hard to illustrate the complex light-matter interaction in the 2D plane, a 3D electric field distribution of optimal NC texture with embedded Ge NP has also been plotted for those four wavelengths in the fourth row of the Fig. \ref{electric_field} grid.

For the planar CdTe TFSCs, the Fabry-Perot resonance mode constructed from the constructive and destructive interference of incoming and reflected lights can be seen in Fig. \ref{electric_field}(a-d) \cite{LigManPho}. Additionally, the thickness and width of this Fabry-Perot resonance mode can be seen increasing with the increase in wavelength. As for optimal NC texture, having a constant periodicity allows it to work as a diffraction grating \cite{LigHarThin}. Broadband absorption through NC texture can be achieved in two ways \cite{NanPhoMan}. The texture layer significantly suppresses the reflective losses at shorter wavelengths and focuses light into the absorber layer as seen in Fig. \ref{electric_field}(e-f). On the other hand, at longer wavelengths, CdTe is less absorptive, and therefore incident light can not be absorbed in a single path length, i.e., path taken by light to travel without being reflected. With the help of the back reflector, the diffraction distribution pattern arises as seen in Fig. \ref{electric_field}(g-h), increasing the optical path length to several folds \cite{OptPerMod}. For the both NC texture and Ge NP configuration, the electric field distribution plot (Fig. \ref{electric_field}(i-j)) is somewhat similar to the only NC texture configuration at 545 and 785 nm owing to the fact that Ge NP contributes less to the light coupling due to its low scattering cross-section at these wavelengths as described in the previous section. However, at long wavelengths i.e. 935 nm and 1035 nm, Ge NP can be found to increase light coupling through its strong forward scattering mechanism and near field effect demonstrated in Fig. \ref{electric_field}(k-l) and also in the 3D field distribution of Fig. \ref{electric_field}(o-p). Since both the mechanisms are radiative processes, they can transfer energy with the help of photons \cite{EffPlasMet}. Several hotspots are visible at 1035 nm (Fig. \ref{electric_field}(l)), suggesting the strong light coupling in the presence of Ge NP can contribute to the electron-hole pair generations and thus increase overall PCE \cite{EnhLigAbsUltra}. Clearly, optimal NC texture and Ge NP configuration together provide an improved method to utilize light to increase carrier generation at both short and long wavelengths compared to the other two configurations.

\subsubsection{P-V \& J-V curve analysis}

Additionally, the current density vs voltage (J-V curve), and output power vs voltage (P-V curve) have been plotted in Fig. \ref{PV_IV}(a-b), respectively. The figures suggest that tapering the top surface of both the CdS-CdTe layer in a NC shape with a base diameter of 600 nm and a height of 400 nm can significantly enhance the current density. Adding a 300 nm diameter Ge NP embedded 500 nm below the top surface can even increase the current density further. Despite the reduction in solar cell voltage in both cases, the solar cell quality, i.e., fill factor, increases, contributing to achieving a higher cell output power compared to the bare CdTe TFSCs as seen in Fig. \ref{PV_IV}(b). Finally, the absorption spectra of bare CdTe TFSCs has been plotted against the optimal structure that is modified with a NC texture and Ge NP. Fig. \ref{PV_IV}(c) shows that the bare flat structure suffers from poor absorption over many of the incident light wavelengths, especially in the NIR region. On the contrary, the optimal structure absorbs almost all the incoming light of solar spectrum AM 1.5G, owing to its excellent anti-reflection and light trapping properties over broadband wavelengths. 
Table \ref{comparison} compares the performance parameters of the proposed structure with that of other recently investigated CdTe TFSCs.

\begin{figure*}[!t]
 \centering
 \includegraphics[width=\textwidth]{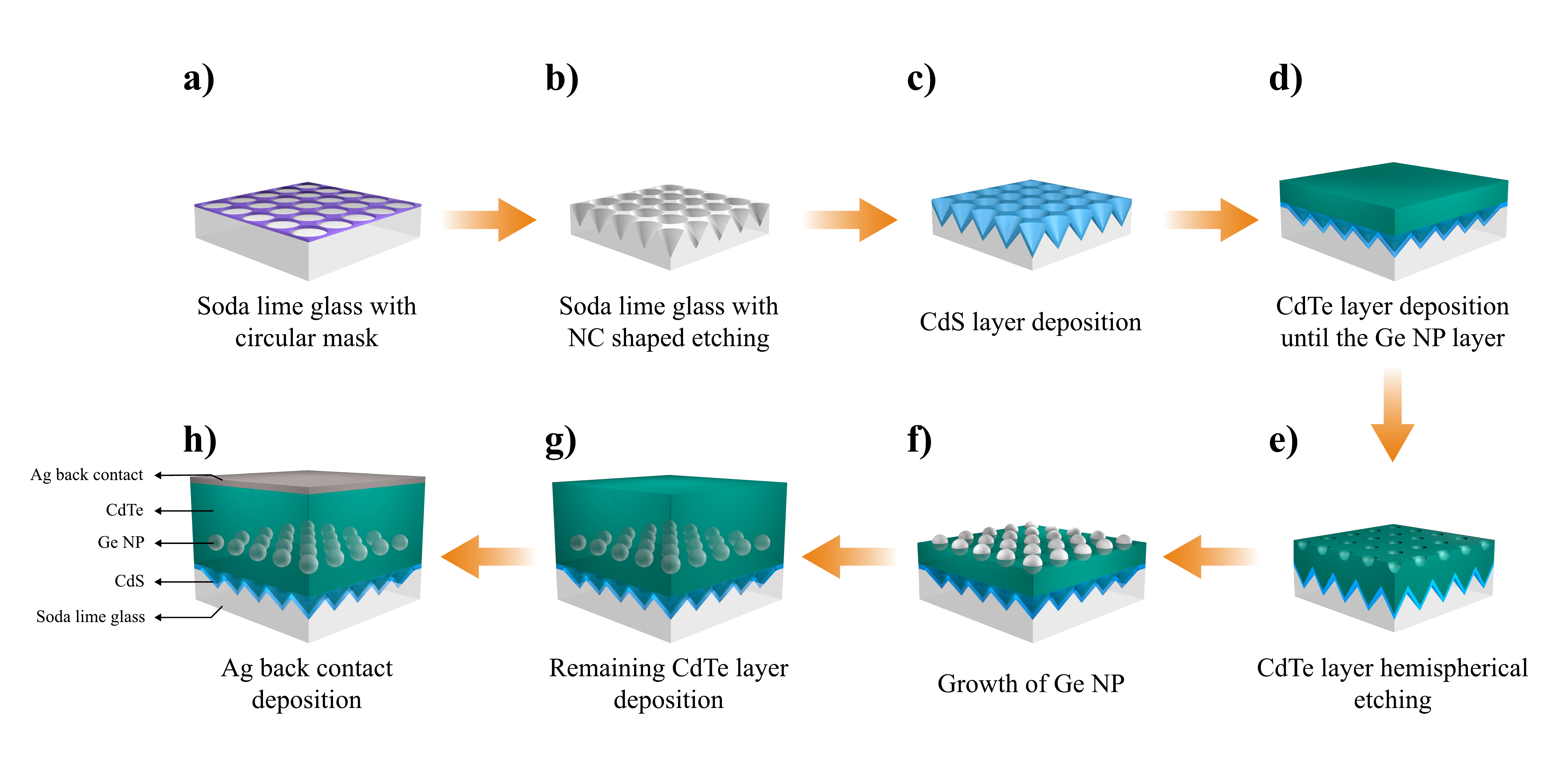}
 \caption{The step-by-step procedure for fabricating the proposed structure, (a) circular SiO\textsubscript{2}
hard mask on soda lime glass, (b) wet etching of inverted NC array using KOH, ethanol, and deionized water solution, (c) CdS layer deposition by closed-spaced sublimation (CSS), (d)
CdTe layer deposition by CSS until the depth of Ge NP formation, (e) array of semicircular inverted pits by electron beam lithography (EBL), (f) growth of Ge NP by molecular beam epitaxy (MBE), (g)
deposition of remaining CdTe absorbing layer, (h) Ag back contact deposition by magnetron
sputtering.}
 \label{fabrication_steps}
\end{figure*}

\subsection{Polarization \& incident angle variation analysis}

Although all the simulations conducted in this study are for TM polarized light in normal incident angle, the actual sunlight is unpolarized light and follows an oblique solar path. This is even more true if the solar cell is being used in a country located far from the equator where the incident angle is further sloped. Therefore, incident and polarization variation studies are critical to gain a better understanding of the phenomenon of light interaction with the NC structures and embedded Ge NPs in the CdTe TFSC. Fig. \ref{polarization_incident_angle}(a) shows the polarization angle variation from 0$^{\circ}$ to 90$^{\circ}$ as a function of incident light wavelengths for the CdTe solar cells with optimal NC texture and embedded Ge NP. It is essential to note that 0$^{\circ}$ corresponds to the TM polarized light whereas 90$^{\circ}$ refers to TE polarized light. Additionally, the absorption remained the same for all polarized angles since both the NC texture and Ge spherical NP have x-y axial symmetry. This means the optimal structure obtained in this study can keep a consistent performance for any polarized light. Next, the incident angle has been varied from 0$^{\circ}$ to 70$^{\circ}$ as a function of incident light wavelengths for the same optimal
structure. The results shown in \ref{polarization_incident_angle}b are consistent with the previously reported study where the nanotextured front surfaces have shown almost no significant efficiency degradation of the solar cell over an incident angle of 0$^{\circ}$ to 70$^{\circ}$ \cite{LigHarThin, NanodomeSolCel}. Although there is a slight absorption degradation at higher incident angles, the average absorption is maintained well above 85\% over all the angle variations as shown in Fig. S4. The optimal NC texture plays an important role in this aspect, providing enhanced absorption over a broad wavelength range and large incident angle variation \cite{OptAbsEnhAmor}. This further supports the robustness of the optimal structure obtained in this study and its insensitivity against polarization and incident angle of the incoming solar radiation.

\subsection{Proposed fabrication technique}

Over the years, the fabrication process CdTe TFSCs have been streamlined industrially \cite{History-of-Photovoltaics}. While optimizing the proposed structure in this study, one of the goals was its ease of fabrication by implementing additional steps to existing processes. Hence, additional methods for the NC structure and embedding of Ge NP have to be adopted to fabricate the proposed structure shown in Fig. \ref{fdtd_charge_setup}. The suggested fabrication process is summarized in Fig. \ref{fabrication_steps}. CdTe TFSCs are generally fabricated in a top-down approach, i.e., CdS (window layer) is deposited on a substrate, then CdTe (absorber layer) and then the back contact \cite{History-of-Photovoltaics, OperationandReliability}.  Similarly, top-down approach was taken by Tavakoli et al., where layers of a perovskite solar cell were deposited on a plastic textured with hexagonally ordered inverted NC structure, essentially the plastic substrate acted as the mold for the NC texture formation \cite{inano-perv}.  This approach can be taken to potentially fabricate the NC textures in the proposed structure. Industrially, soda lime glass is used as a substrate for CdTe TFSC production due to its minimal parasitic light absorption and cost effectiveness \cite{History-of-Photovoltaics}.

Ding et al. devised a method to imprint high resolution images based on inverted nanopyramid textures on Si wafers and also claimed the method can be applied to produce nanowires and nanoholes. \cite{img-patterned-Si}. This method can potentially be adopted to imprint square arrays of inverted NCs by preparing the CAD file containing the NC pattern to produce the SiO\textsubscript{2} hard mask directly on a soda-lime substrate using laser ablation. Subsequently, the NCs can be produced using wet etching solution comprising potassium hydroxide (KOH), ethanol, and deionized water. Additionally, potassium hydroxide was used by Sakai et al. as an etchant to produce microstructures on soda-lime \cite{chem-etch-slime}. So, this method of NC texturing may be potentially used on soda lime substrate as well. To deposit the CdS and CdTe layers, chemical bath deposition (CBD) and closed space sublimation (CSS) can be used, respectively \cite{History-of-Photovoltaics}. Ruiz-Ortega et al. fabricated CdS ultrathin films with thickness ranging from 27-48 nm using CSS, hence the CdS window layer thickness of 100 nm for the proposed configuration is potentially achievable \cite{AnalysisofCdS}.

For the next part, deposition of CdTe layer has to be done in two steps for enabling placement of a Ge NP array. This could be thought of as a layer-by-layer approach where CdTe filled up to a thickness, then fabrication and placement of Ge NP, then again deposition of CdTe as depicted in Fig. \ref{fabrication_steps}(d-g). This approach was taken for embedding Ag NPs in amorphous Si by Santbergen et al. and also an almost similar approach can be seen for embedding SiO\textsubscript{2} NPs in organic solar cells where a mixture of PEDOT:PSS and SiO\textsubscript{2} NPs was spin coated on to a substrate and then P3HT:PC61BM was deposited on top \cite{silver-np-si, organ-np}. Henceforth, CdTe can potentially be deposited using CSS by maintaining substrate temperature at 500$^{\circ}$C up to a thickness of 400 nm so that Ge NP can be placed \cite{History-of-Photovoltaics}. Bollani et al. reported a procedure to embed ordered arrays of Ga NPs on the surface of Si substrate, where pit arrays acting as place holders were patterned on to the Si using electron beam lithography (EBL) and the Ga NPs were grown using molecular beam epitaxy (MBE) in the ordered pits \cite{orderedGa}. This method could potentially be expanded to CdTe, where the pit arrays could be textured on to CdTe using EBL and the Ge NPs could be produced using molecular beam epitaxy following the methodology of Aouassa et al., where they deposited Mn doped Ge NPs on to SiO\textsubscript{2} thin-films for solar cell applications \cite{mngermaniumnp}.

Once Ge NPs are implanted, CdTe deposition can be continued for another 800 nm to attain a final thickness of 1600 nm of the absorber layer. Moving forward the structure is to be annealed in the presence of Cl atmosphere also known as Cl treatment or activation treatment, where Cl is supplied by CdCl\textsubscript{2} or alternatively MgCl\textsubscript{2} which is non-toxic and cheaper \cite{History-of-Photovoltaics, OperationandReliability}.  This ensures the recrystallization of the CdTe removing lattice defects from the CdS/CdTe heterojunction interface and reduce lattice mismatch \cite{History-of-Photovoltaics}. This step can be followed by a chemical treatment of the CdTe surface using Br\textsubscript{2}-methanol or a mixture of HNO\textsubscript{3}/HPO\textsubscript{3} acids to create a Te-rich layer. This step has the potential to  facilitate alignment between the valence band of CdTe and the work function of the metal contact. \cite{History-of-Photovoltaics}. Lastly the Ag back contact can be deposited using magnetron sputtering following the methodology proposed by Eze et al \cite{bcsputter}.

\section*{Conclusions}
In conclusion, the results presented in this computational study reveal that a composite light-trapping technique combining NC-shaped surface texturing and embedded Ge NPs can significantly enhance light absorption in CdTe TFSCs, achieving 95\% absorption compared to 64\% in CdTe TFSCs without any light-trapping approach. The NC texture, optimized with a 600 nm base diameter and a 400 nm height, effectively reduces surface reflection across the incident light spectrum. Additionally, it acts as a diffraction grating, directing light into the absorber layer to increase carrier generation by extending optical path lengths. Embedding 300 nm Ge NPs at a depth of 500 nm from the top surface helps address weak absorption of CdTe at near-infrared wavelengths. Scattering cross-sections calculated using Mie theory reveal that, upon light incidence, the Ge NPs exhibit strong forward-scattering behavior, coupling scattered light deeper into the CdTe absorber and enhancing current density. The optimal NC texture with Ge NPs yields the highest simulated values for short-circuit current density (J\textsubscript{sc}) at 35.38 mA/cm\textsuperscript{2}, open-circuit voltage (V\textsubscript{oc}) at 933.2 mV, fill factor at 0.8406, and power conversion efficiency (PCE) at 27.76\%. Furthermore, additional analyses indicate that the performance of this structure is stable under varying polarization angles (0$^{\circ}$ to 90$^{\circ}$) and incident angles (0$^{\circ}$ to 70$^{\circ}$) of the incident solar radiation, demonstrating its robustness for practical applications. This resilience can potentially make the optimized CdTe TFSCs viable for integration into miniaturized electronic devices. Although the optimal solar cell structure has not been fabricated in this computational study, a fabrication process has been proposed connecting the established fabrication-related pieces of literature. Future work will focus on actual device fabrication to establish performance benchmarks, aiming for a cost-effective approach without compromising efficiency.

\section*{Conflicts of interest}
The authors declare that they have no known competing financial interests or personal relationships that could have appeared to influence the work reported in this paper.

\section*{Data availability}

Data will be made available on request.

\section*{Acknowledgements}

The authors would like to thank Independent University, Bangladesh (IUB) for providing funds and other logistical support for this research.

%%%END OF MAIN TEXT%%%


\balance

\renewcommand\refname{References}

%%%REFERENCES%%%
\bibliography{rsc} %You need to replace "rsc" on this line with the name of your .bib file

\providecommand*{\mcitethebibliography}{\thebibliography}
\csname @ifundefined\endcsname{endmcitethebibliography}
{\let\endmcitethebibliography\endthebibliography}{}
\begin{mcitethebibliography}{82}
\providecommand*{\natexlab}[1]{#1}
\providecommand*{\mciteSetBstSublistMode}[1]{}
\providecommand*{\mciteSetBstMaxWidthForm}[2]{}
\providecommand*{\mciteBstWouldAddEndPuncttrue}
  {\def\EndOfBibitem{\unskip.}}
\providecommand*{\mciteBstWouldAddEndPunctfalse}
  {\let\EndOfBibitem\relax}
\providecommand*{\mciteSetBstMidEndSepPunct}[3]{}
\providecommand*{\mciteSetBstSublistLabelBeginEnd}[3]{}
\providecommand*{\EndOfBibitem}{}
\mciteSetBstSublistMode{f}
\mciteSetBstMaxWidthForm{subitem}
{(\emph{\alph{mcitesubitemcount}})}
\mciteSetBstSublistLabelBeginEnd{\mcitemaxwidthsubitemform\space}
{\relax}{\relax}

\bibitem[Maka and Alabid(2022)]{SolEneTec}
A.~O. Maka and J.~M. Alabid, \emph{Clean Energy}, 2022, \textbf{6}, 476--483\relax
\mciteBstWouldAddEndPuncttrue
\mciteSetBstMidEndSepPunct{\mcitedefaultmidpunct}
{\mcitedefaultendpunct}{\mcitedefaultseppunct}\relax
\EndOfBibitem
\bibitem[IEA()]{IEA-Ren}
\url{https://www.iea.org/reports/renewables-2024/executive-summary}\relax
\mciteBstWouldAddEndPuncttrue
\mciteSetBstMidEndSepPunct{\mcitedefaultmidpunct}
{\mcitedefaultendpunct}{\mcitedefaultseppunct}\relax
\EndOfBibitem
\bibitem[IEA()]{IEA-Wor}
\url{https://www.iea.org/reports/world-energy-outlook-2023/executive-summary}\relax
\mciteBstWouldAddEndPuncttrue
\mciteSetBstMidEndSepPunct{\mcitedefaultmidpunct}
{\mcitedefaultendpunct}{\mcitedefaultseppunct}\relax
\EndOfBibitem
\bibitem[Regmi and Subramaniam(2021)]{IntPhoAlt}
G.~Regmi and V.~Subramaniam, \emph{Sustainable Material Solutions for Solar Energy Technologies}, Elsevier, 2021, pp. 131--173\relax
\mciteBstWouldAddEndPuncttrue
\mciteSetBstMidEndSepPunct{\mcitedefaultmidpunct}
{\mcitedefaultendpunct}{\mcitedefaultseppunct}\relax
\EndOfBibitem
\bibitem[Hashmi \emph{et~al.}(2023)Hashmi, Hossain, and Imtiaz]{EleOptPara}
G.~Hashmi, M.~S. Hossain and M.~H. Imtiaz, \emph{Journal of Theoretical and Applied Physics}, 2023, \textbf{17}, 172328--172329\relax
\mciteBstWouldAddEndPuncttrue
\mciteSetBstMidEndSepPunct{\mcitedefaultmidpunct}
{\mcitedefaultendpunct}{\mcitedefaultseppunct}\relax
\EndOfBibitem
\bibitem[Bosio \emph{et~al.}(2020)Bosio, Pasini, and Romeo]{TheHisPho}
A.~Bosio, S.~Pasini and N.~Romeo, \emph{Coatings}, 2020, \textbf{10}, 344\relax
\mciteBstWouldAddEndPuncttrue
\mciteSetBstMidEndSepPunct{\mcitedefaultmidpunct}
{\mcitedefaultendpunct}{\mcitedefaultseppunct}\relax
\EndOfBibitem
\bibitem[{\c{C}}etinkaya \emph{et~al.}(2022){\c{C}}etinkaya, {\c{C}}okduygulular, K{\i}nac{\i}, G{\"u}zel{\c{c}}imen, {\"O}zen, S{\"o}nmez, and {\"O}z{\c{c}}elik]{HigImpLig}
{\c{C}}.~{\c{C}}etinkaya, E.~{\c{C}}okduygulular, B.~K{\i}nac{\i}, F.~G{\"u}zel{\c{c}}imen, Y.~{\"O}zen, N.~A. S{\"o}nmez and S.~{\"O}z{\c{c}}elik, \emph{Scientific Reports}, 2022, \textbf{12}, 11245\relax
\mciteBstWouldAddEndPuncttrue
\mciteSetBstMidEndSepPunct{\mcitedefaultmidpunct}
{\mcitedefaultendpunct}{\mcitedefaultseppunct}\relax
\EndOfBibitem
\bibitem[US-()]{US-DOE}
\url{https://www.energy.gov/eere/solar/cadmium-telluride}\relax
\mciteBstWouldAddEndPuncttrue
\mciteSetBstMidEndSepPunct{\mcitedefaultmidpunct}
{\mcitedefaultendpunct}{\mcitedefaultseppunct}\relax
\EndOfBibitem
\bibitem[Ali \emph{et~al.}(2022)Ali, El-Mellouhi, Mitra, and A{\"\i}ssa]{ResProPlas}
A.~Ali, F.~El-Mellouhi, A.~Mitra and B.~A{\"\i}ssa, \emph{Nanomaterials}, 2022, \textbf{12}, 788\relax
\mciteBstWouldAddEndPuncttrue
\mciteSetBstMidEndSepPunct{\mcitedefaultmidpunct}
{\mcitedefaultendpunct}{\mcitedefaultseppunct}\relax
\EndOfBibitem
\bibitem[Wu \emph{et~al.}(2020)Wu, Wang, Batmunkh, Bati, Yang, Jiang, Hou, Shapter, and Priya]{MulNanMat}
C.~Wu, K.~Wang, M.~Batmunkh, A.~S. Bati, D.~Yang, Y.~Jiang, Y.~Hou, J.~G. Shapter and S.~Priya, \emph{Nano Energy}, 2020, \textbf{70}, 104480\relax
\mciteBstWouldAddEndPuncttrue
\mciteSetBstMidEndSepPunct{\mcitedefaultmidpunct}
{\mcitedefaultendpunct}{\mcitedefaultseppunct}\relax
\EndOfBibitem
\bibitem[Starczewska and K{\k{e}}pi{\'n}ska(2024)]{PhoCryStr}
A.~Starczewska and M.~K{\k{e}}pi{\'n}ska, \emph{Materials}, 2024, \textbf{17}, 1196\relax
\mciteBstWouldAddEndPuncttrue
\mciteSetBstMidEndSepPunct{\mcitedefaultmidpunct}
{\mcitedefaultendpunct}{\mcitedefaultseppunct}\relax
\EndOfBibitem
\bibitem[Noor \emph{et~al.}(2023)Noor, Hossain, Pramanik, Suny, Sultan, Tohfa, and Chowdhury]{STI_CNT}
T.~Noor, M.~H. Hossain, A.~Pramanik, A.~A. Suny, R.~B. Sultan, S.~Tohfa and M.~H. Chowdhury, 2023 5th International Conference on Sustainable Technologies for Industry 5.0 (STI), 2023, pp. 1--6\relax
\mciteBstWouldAddEndPuncttrue
\mciteSetBstMidEndSepPunct{\mcitedefaultmidpunct}
{\mcitedefaultendpunct}{\mcitedefaultseppunct}\relax
\EndOfBibitem
\bibitem[Kim \emph{et~al.}(2020)Kim, Lee, and Kwak]{SurTexMet}
M.~S. Kim, J.~H. Lee and M.~K. Kwak, \emph{International Journal of Precision Engineering and Manufacturing}, 2020, \textbf{21}, 1389--1398\relax
\mciteBstWouldAddEndPuncttrue
\mciteSetBstMidEndSepPunct{\mcitedefaultmidpunct}
{\mcitedefaultendpunct}{\mcitedefaultseppunct}\relax
\EndOfBibitem
\bibitem[Suny \emph{et~al.}(2023)Suny, Tohfa, Sultan, Hossain, Noor, and Chowdhury]{STI_ST}
A.~A. Suny, S.~Tohfa, R.~B. Sultan, M.~H. Hossain, T.~Noor and M.~H. Chowdhury, 2023 5th International Conference on Sustainable Technologies for Industry 5.0 (STI), 2023, pp. 1--6\relax
\mciteBstWouldAddEndPuncttrue
\mciteSetBstMidEndSepPunct{\mcitedefaultmidpunct}
{\mcitedefaultendpunct}{\mcitedefaultseppunct}\relax
\EndOfBibitem
\bibitem[Hou \emph{et~al.}(2022)Hou, Zhang, Zheng, Lu, Pogrebnyakov, Wu, Yoon, Yang, Zheng, Gopalan,\emph{et~al.}]{HomOptFie}
Y.~Hou, J.~Zhang, X.~Zheng, Y.~Lu, A.~Pogrebnyakov, H.~Wu, J.~Yoon, D.~Yang, L.~Zheng, V.~Gopalan \emph{et~al.}, \emph{ACS Energy Letters}, 2022, \textbf{7}, 1657--1671\relax
\mciteBstWouldAddEndPuncttrue
\mciteSetBstMidEndSepPunct{\mcitedefaultmidpunct}
{\mcitedefaultendpunct}{\mcitedefaultseppunct}\relax
\EndOfBibitem
\bibitem[Abu-Elmaaty \emph{et~al.}(2024)Abu-Elmaaty, Ismail, Sabeeh, and Khawaji]{OptThinSi}
B.~E. Abu-Elmaaty, T.~Ismail, A.~H. Sabeeh and I.~H. Khawaji, \emph{Applied Optics}, 2024, \textbf{63}, 3885--3891\relax
\mciteBstWouldAddEndPuncttrue
\mciteSetBstMidEndSepPunct{\mcitedefaultmidpunct}
{\mcitedefaultendpunct}{\mcitedefaultseppunct}\relax
\EndOfBibitem
\bibitem[Li \emph{et~al.}(2020)Li, Hu, Yang, and Zhu]{TheInvBro}
H.~Li, Y.~Hu, Y.~Yang and Y.~Zhu, \emph{Solar Energy Materials and Solar Cells}, 2020, \textbf{211}, 110529\relax
\mciteBstWouldAddEndPuncttrue
\mciteSetBstMidEndSepPunct{\mcitedefaultmidpunct}
{\mcitedefaultendpunct}{\mcitedefaultseppunct}\relax
\EndOfBibitem
\bibitem[Pritom \emph{et~al.}(2023)Pritom, Sikder, Zaman, and Hossain]{PlaEnhPara}
Y.~A. Pritom, D.~K. Sikder, S.~Zaman and M.~Hossain, \emph{Nanoscale Advances}, 2023, \textbf{5}, 4986--4995\relax
\mciteBstWouldAddEndPuncttrue
\mciteSetBstMidEndSepPunct{\mcitedefaultmidpunct}
{\mcitedefaultendpunct}{\mcitedefaultseppunct}\relax
\EndOfBibitem
\bibitem[Nagel and Scarpulla(2013)]{EnhLigAbsThin}
J.~R. Nagel and M.~A. Scarpulla, \emph{Applied Physics Letters}, 2013, \textbf{102}, 151111\relax
\mciteBstWouldAddEndPuncttrue
\mciteSetBstMidEndSepPunct{\mcitedefaultmidpunct}
{\mcitedefaultendpunct}{\mcitedefaultseppunct}\relax
\EndOfBibitem
\bibitem[Yin \emph{et~al.}(2014)Yin, Yu, Liu, Ye, Zhang, Cui, Yu, Wang, and Zhang]{DesPlasSol}
Y.~Yin, Z.~Yu, Y.~Liu, H.~Ye, W.~Zhang, Q.~Cui, X.~Yu, P.~Wang and Y.~Zhang, \emph{Optics Communications}, 2014, \textbf{333}, 213--218\relax
\mciteBstWouldAddEndPuncttrue
\mciteSetBstMidEndSepPunct{\mcitedefaultmidpunct}
{\mcitedefaultendpunct}{\mcitedefaultseppunct}\relax
\EndOfBibitem
\bibitem[Wang \emph{et~al.}(2014)Wang, Shao, Zhang, Zhuge, Wu, and Zhang]{BroOmnAnt}
Y.~Wang, B.~Shao, Z.~Zhang, L.~Zhuge, X.~Wu and R.~Zhang, \emph{Optics Communications}, 2014, \textbf{316}, 37--41\relax
\mciteBstWouldAddEndPuncttrue
\mciteSetBstMidEndSepPunct{\mcitedefaultmidpunct}
{\mcitedefaultendpunct}{\mcitedefaultseppunct}\relax
\EndOfBibitem
\bibitem[Zhu \emph{et~al.}(2010)Zhu, Yu, Fan, and Cui]{NanPhoMan}
J.~Zhu, Z.~Yu, S.~Fan and Y.~Cui, \emph{Materials Science and Engineering: R: Reports}, 2010, \textbf{70}, 330--340\relax
\mciteBstWouldAddEndPuncttrue
\mciteSetBstMidEndSepPunct{\mcitedefaultmidpunct}
{\mcitedefaultendpunct}{\mcitedefaultseppunct}\relax
\EndOfBibitem
\bibitem[Smay \emph{et~al.}(2018)Smay, Rashwan, and Then]{OptNanPara}
J.~Smay, O.~Rashwan and J.~Then, ASME International Mechanical Engineering Congress and Exposition, 2018, p. V06BT08A025\relax
\mciteBstWouldAddEndPuncttrue
\mciteSetBstMidEndSepPunct{\mcitedefaultmidpunct}
{\mcitedefaultendpunct}{\mcitedefaultseppunct}\relax
\EndOfBibitem
\bibitem[Vadavalli \emph{et~al.}(2014)Vadavalli, Valligatla, Neelamraju, Dar, Chiasera, Ferrari, and Desai]{OptProGer}
S.~Vadavalli, S.~Valligatla, B.~Neelamraju, M.~H. Dar, A.~Chiasera, M.~Ferrari and N.~R. Desai, \emph{Frontiers in Physics}, 2014, \textbf{2}, 57\relax
\mciteBstWouldAddEndPuncttrue
\mciteSetBstMidEndSepPunct{\mcitedefaultmidpunct}
{\mcitedefaultendpunct}{\mcitedefaultseppunct}\relax
\EndOfBibitem
\bibitem[Brongersma \emph{et~al.}(2014)Brongersma, Cui, and Fan]{LigManPho}
M.~L. Brongersma, Y.~Cui and S.~Fan, \emph{Nature materials}, 2014, \textbf{13}, 451--460\relax
\mciteBstWouldAddEndPuncttrue
\mciteSetBstMidEndSepPunct{\mcitedefaultmidpunct}
{\mcitedefaultendpunct}{\mcitedefaultseppunct}\relax
\EndOfBibitem
\bibitem[Zhang \emph{et~al.}(2019)Zhang, Li, Deng, Yan, Wang, Chen, Sun, and Huang]{EnhPhoPer}
C.~Zhang, Z.~Li, X.~Deng, B.~Yan, Z.~Wang, X.~Chen, Z.~Sun and S.~Huang, \emph{Solar Energy}, 2019, \textbf{188}, 839--848\relax
\mciteBstWouldAddEndPuncttrue
\mciteSetBstMidEndSepPunct{\mcitedefaultmidpunct}
{\mcitedefaultendpunct}{\mcitedefaultseppunct}\relax
\EndOfBibitem
\bibitem[Ma \emph{et~al.}(2017)Ma, Yan, Huang, and Yang]{DirScaGer}
C.~Ma, J.~Yan, Y.~Huang and G.~Yang, \emph{Advanced Optical Materials}, 2017, \textbf{5}, 1700761\relax
\mciteBstWouldAddEndPuncttrue
\mciteSetBstMidEndSepPunct{\mcitedefaultmidpunct}
{\mcitedefaultendpunct}{\mcitedefaultseppunct}\relax
\EndOfBibitem
\bibitem[Naik \emph{et~al.}(2013)Naik, Shalaev, and Boltasseva]{AltPlasMat}
G.~V. Naik, V.~M. Shalaev and A.~Boltasseva, \emph{Advanced materials}, 2013, \textbf{25}, 3264--3294\relax
\mciteBstWouldAddEndPuncttrue
\mciteSetBstMidEndSepPunct{\mcitedefaultmidpunct}
{\mcitedefaultendpunct}{\mcitedefaultseppunct}\relax
\EndOfBibitem
\bibitem[Boltasseva and Atwater(2011)]{LowPlasMet}
A.~Boltasseva and H.~A. Atwater, \emph{Science}, 2011, \textbf{331}, 290--291\relax
\mciteBstWouldAddEndPuncttrue
\mciteSetBstMidEndSepPunct{\mcitedefaultmidpunct}
{\mcitedefaultendpunct}{\mcitedefaultseppunct}\relax
\EndOfBibitem
\bibitem[Thakore \emph{et~al.}(2019)Thakore, Tang, Conley, Ala-Nissila, and Karttunen]{TheResSem}
V.~Thakore, J.~Tang, K.~Conley, T.~Ala-Nissila and M.~Karttunen, \emph{Advanced Theory and Simulations}, 2019, \textbf{2}, 1800100\relax
\mciteBstWouldAddEndPuncttrue
\mciteSetBstMidEndSepPunct{\mcitedefaultmidpunct}
{\mcitedefaultendpunct}{\mcitedefaultseppunct}\relax
\EndOfBibitem
\bibitem[Ishii \emph{et~al.}(2017)Ishii, Chen, Okuyama, and Nagao]{ResOptAbs}
S.~Ishii, K.~Chen, H.~Okuyama and T.~Nagao, \emph{Advanced Optical Materials}, 2017, \textbf{5}, 1600902\relax
\mciteBstWouldAddEndPuncttrue
\mciteSetBstMidEndSepPunct{\mcitedefaultmidpunct}
{\mcitedefaultendpunct}{\mcitedefaultseppunct}\relax
\EndOfBibitem
\bibitem[Luo \emph{et~al.}(2022)Luo, Wang, Zhu, Zhu, and Li]{DirLigSca}
S.~Luo, S.~Wang, Y.~Zhu, E.~Zhu and Z.~Li, \emph{Optics Communications}, 2022, \textbf{507}, 127606\relax
\mciteBstWouldAddEndPuncttrue
\mciteSetBstMidEndSepPunct{\mcitedefaultmidpunct}
{\mcitedefaultendpunct}{\mcitedefaultseppunct}\relax
\EndOfBibitem
\bibitem[Carolan(2017)]{RecAdvGer}
D.~Carolan, \emph{Progress in materials science}, 2017, \textbf{90}, 128--158\relax
\mciteBstWouldAddEndPuncttrue
\mciteSetBstMidEndSepPunct{\mcitedefaultmidpunct}
{\mcitedefaultendpunct}{\mcitedefaultseppunct}\relax
\EndOfBibitem
\bibitem[Palik(1998)]{HanOptCon}
E.~Palik, \emph{Handbook of Optical Constants of Solids}, Elsevier Science, 1998\relax
\mciteBstWouldAddEndPuncttrue
\mciteSetBstMidEndSepPunct{\mcitedefaultmidpunct}
{\mcitedefaultendpunct}{\mcitedefaultseppunct}\relax
\EndOfBibitem
\bibitem[Chen \emph{et~al.}(2023)Chen, Liu, Liu, Mao, Wei, Ji, Yang, Bao, Yang, and Wang]{AbsEnhGaAs}
X.~Chen, Q.~Liu, W.~Liu, X.~Mao, B.~Wei, C.~Ji, G.~Yang, Y.~Bao, F.~Yang and X.~Wang, \emph{Applied Optics}, 2023, \textbf{62}, 7111--7118\relax
\mciteBstWouldAddEndPuncttrue
\mciteSetBstMidEndSepPunct{\mcitedefaultmidpunct}
{\mcitedefaultendpunct}{\mcitedefaultseppunct}\relax
\EndOfBibitem
\bibitem[Treharne \emph{et~al.}(2011)Treharne, Seymour-Pierce, Durose, Hutchings, Roncallo, and Lane]{OptDesFab}
R.~Treharne, A.~Seymour-Pierce, K.~Durose, K.~Hutchings, S.~Roncallo and D.~Lane, Journal of Physics: Conference Series, 2011, p. 012038\relax
\mciteBstWouldAddEndPuncttrue
\mciteSetBstMidEndSepPunct{\mcitedefaultmidpunct}
{\mcitedefaultendpunct}{\mcitedefaultseppunct}\relax
\EndOfBibitem
\bibitem[Hagemann \emph{et~al.}(1975)Hagemann, Gudat, and Kunz]{OptConFar}
H.-J. Hagemann, W.~Gudat and C.~Kunz, \emph{JOSA}, 1975, \textbf{65}, 742--744\relax
\mciteBstWouldAddEndPuncttrue
\mciteSetBstMidEndSepPunct{\mcitedefaultmidpunct}
{\mcitedefaultendpunct}{\mcitedefaultseppunct}\relax
\EndOfBibitem
\bibitem[Rangel-C{\'a}rdenas and Sobral(2017)]{OptAbsEnhCdTe}
J.~Rangel-C{\'a}rdenas and H.~Sobral, \emph{Materials}, 2017, \textbf{10}, 607\relax
\mciteBstWouldAddEndPuncttrue
\mciteSetBstMidEndSepPunct{\mcitedefaultmidpunct}
{\mcitedefaultendpunct}{\mcitedefaultseppunct}\relax
\EndOfBibitem
\bibitem[Fardi and Buny(2013)]{ChaModCdSCdTe}
H.~Fardi and F.~Buny, \emph{International Journal of Photoenergy}, 2013, \textbf{2013}, 576952\relax
\mciteBstWouldAddEndPuncttrue
\mciteSetBstMidEndSepPunct{\mcitedefaultmidpunct}
{\mcitedefaultendpunct}{\mcitedefaultseppunct}\relax
\EndOfBibitem
\bibitem[Levinson \emph{et~al.}(2005)Levinson, Berdahl, and Akbari]{absorbance}
R.~Levinson, P.~Berdahl and H.~Akbari, \emph{Solar energy materials and solar cells}, 2005, \textbf{89}, 319--349\relax
\mciteBstWouldAddEndPuncttrue
\mciteSetBstMidEndSepPunct{\mcitedefaultmidpunct}
{\mcitedefaultendpunct}{\mcitedefaultseppunct}\relax
\EndOfBibitem
\bibitem[Su \emph{et~al.}(2021)Su, Yang, Xu, Tang, Yi, Zheng, Zhao, Liu, Wu, and Li]{num_analysis}
J.~Su, H.~Yang, Y.~Xu, Y.~Tang, Z.~Yi, F.~Zheng, F.~Zhao, L.~Liu, P.~Wu and H.~Li, \emph{Coatings}, 2021, \textbf{11}, 748\relax
\mciteBstWouldAddEndPuncttrue
\mciteSetBstMidEndSepPunct{\mcitedefaultmidpunct}
{\mcitedefaultendpunct}{\mcitedefaultseppunct}\relax
\EndOfBibitem
\bibitem[ans()]{ansys}
\url{https://optics.ansys.com/hc/en-us/articles/360042165634-Solar-cell-methodology}\relax
\mciteBstWouldAddEndPuncttrue
\mciteSetBstMidEndSepPunct{\mcitedefaultmidpunct}
{\mcitedefaultendpunct}{\mcitedefaultseppunct}\relax
\EndOfBibitem
\bibitem[AM1()]{AM1.5}
\url{https://www.nrel.gov/grid/solar-resource/spectra-am1.5.html}\relax
\mciteBstWouldAddEndPuncttrue
\mciteSetBstMidEndSepPunct{\mcitedefaultmidpunct}
{\mcitedefaultendpunct}{\mcitedefaultseppunct}\relax
\EndOfBibitem
\bibitem[Liu \emph{et~al.}(2022)Liu, Du, Yin, Bai, and Liu]{MicNanoLig}
H.~Liu, Y.~Du, X.~Yin, M.~Bai and W.~Liu, \emph{Journal of Nanomaterials}, 2022, \textbf{2022}, 8139174\relax
\mciteBstWouldAddEndPuncttrue
\mciteSetBstMidEndSepPunct{\mcitedefaultmidpunct}
{\mcitedefaultendpunct}{\mcitedefaultseppunct}\relax
\EndOfBibitem
\bibitem[Ahn \emph{et~al.}(2021)Ahn, Park, Cho, Park, Park, Lee, Hong, Bong, and Yi]{ReaGlaSur}
S.~Ahn, H.~Park, J.~Cho, C.~Park, J.~Park, H.~Lee, K.~Hong, S.~Bong and J.~Yi, \emph{Optik}, 2021, \textbf{229}, 166304\relax
\mciteBstWouldAddEndPuncttrue
\mciteSetBstMidEndSepPunct{\mcitedefaultmidpunct}
{\mcitedefaultendpunct}{\mcitedefaultseppunct}\relax
\EndOfBibitem
\bibitem[Wang \emph{et~al.}(2012)Wang, Yu, Liu, Cui, and Fan]{AbsEnhUltra}
K.~X. Wang, Z.~Yu, V.~Liu, Y.~Cui and S.~Fan, \emph{Nano letters}, 2012, \textbf{12}, 1616--1619\relax
\mciteBstWouldAddEndPuncttrue
\mciteSetBstMidEndSepPunct{\mcitedefaultmidpunct}
{\mcitedefaultendpunct}{\mcitedefaultseppunct}\relax
\EndOfBibitem
\bibitem[Olaimat \emph{et~al.}(2021)Olaimat, Yousefi, and Ramahi]{UsiPlasNano}
M.~M. Olaimat, L.~Yousefi and O.~M. Ramahi, \emph{JOSA B}, 2021, \textbf{38}, 638--651\relax
\mciteBstWouldAddEndPuncttrue
\mciteSetBstMidEndSepPunct{\mcitedefaultmidpunct}
{\mcitedefaultendpunct}{\mcitedefaultseppunct}\relax
\EndOfBibitem
\bibitem[Yu \emph{et~al.}(2010)Yu, Raman, and Fan]{FunLimLig}
Z.~Yu, A.~Raman and S.~Fan, \emph{Optics express}, 2010, \textbf{18}, A366--A380\relax
\mciteBstWouldAddEndPuncttrue
\mciteSetBstMidEndSepPunct{\mcitedefaultmidpunct}
{\mcitedefaultendpunct}{\mcitedefaultseppunct}\relax
\EndOfBibitem
\bibitem[Yu \emph{et~al.}(2010)Yu, Raman, and Fan]{FunLimNano}
Z.~Yu, A.~Raman and S.~Fan, \emph{Proceedings of the National Academy of Sciences}, 2010, \textbf{107}, 17491--17496\relax
\mciteBstWouldAddEndPuncttrue
\mciteSetBstMidEndSepPunct{\mcitedefaultmidpunct}
{\mcitedefaultendpunct}{\mcitedefaultseppunct}\relax
\EndOfBibitem
\bibitem[Shi \emph{et~al.}(2014)Shi, Wang, Liu, Yang, Ma, and Yang]{NanoRearAg}
Y.~Shi, X.~Wang, W.~Liu, T.~Yang, J.~Ma and F.~Yang, \emph{Optics Express}, 2014, \textbf{22}, 20473--20480\relax
\mciteBstWouldAddEndPuncttrue
\mciteSetBstMidEndSepPunct{\mcitedefaultmidpunct}
{\mcitedefaultendpunct}{\mcitedefaultseppunct}\relax
\EndOfBibitem
\bibitem[Mokkapati and Catchpole(2012)]{NanoLigTrap}
S.~Mokkapati and K.~R. Catchpole, \emph{Journal of Applied Physics}, 2012, \textbf{112}, 101101\relax
\mciteBstWouldAddEndPuncttrue
\mciteSetBstMidEndSepPunct{\mcitedefaultmidpunct}
{\mcitedefaultendpunct}{\mcitedefaultseppunct}\relax
\EndOfBibitem
\bibitem[Sheng \emph{et~al.}(1983)Sheng, Bloch, and Stepleman]{WavSelAbs}
P.~Sheng, A.~Bloch and R.~Stepleman, \emph{Applied Physics Letters}, 1983, \textbf{43}, 579--581\relax
\mciteBstWouldAddEndPuncttrue
\mciteSetBstMidEndSepPunct{\mcitedefaultmidpunct}
{\mcitedefaultendpunct}{\mcitedefaultseppunct}\relax
\EndOfBibitem
\bibitem[Zhang \emph{et~al.}(2023)Zhang, Liu, Su, Zhou, Kong, Luo, Gao, Xiong, and Duan]{InvDesPol}
Q.~Zhang, D.~Liu, J.~Su, S.~Zhou, Y.~Kong, H.~Luo, L.~Gao, Y.~Xiong and W.~Duan, \emph{Results in Physics}, 2023, \textbf{45}, 106238\relax
\mciteBstWouldAddEndPuncttrue
\mciteSetBstMidEndSepPunct{\mcitedefaultmidpunct}
{\mcitedefaultendpunct}{\mcitedefaultseppunct}\relax
\EndOfBibitem
\bibitem[Elimelech \emph{et~al.}(1995)Elimelech, Gregory, Jia, and Williams]{ExpTecAgg}
M.~Elimelech, J.~Gregory, X.~Jia and R.~Williams, \emph{Particle Deposition I\& Aggregation}, Butterworth-Heinemann, Woburn, 1995, pp. 263--289\relax
\mciteBstWouldAddEndPuncttrue
\mciteSetBstMidEndSepPunct{\mcitedefaultmidpunct}
{\mcitedefaultendpunct}{\mcitedefaultseppunct}\relax
\EndOfBibitem
\bibitem[Ruffino(2021)]{LigScaCal}
F.~Ruffino, \emph{Micromachines}, 2021, \textbf{12}, 1050\relax
\mciteBstWouldAddEndPuncttrue
\mciteSetBstMidEndSepPunct{\mcitedefaultmidpunct}
{\mcitedefaultendpunct}{\mcitedefaultseppunct}\relax
\EndOfBibitem
\bibitem[Das and Dhawan(2023)]{PlasEnhPho}
P.~K. Das and A.~Dhawan, \emph{RSC advances}, 2023, \textbf{13}, 26780--26792\relax
\mciteBstWouldAddEndPuncttrue
\mciteSetBstMidEndSepPunct{\mcitedefaultmidpunct}
{\mcitedefaultendpunct}{\mcitedefaultseppunct}\relax
\EndOfBibitem
\bibitem[Shi \emph{et~al.}(2017)Shi, Wang, Yu, Yang, and Xu]{EnhOptAbs}
B.~Shi, W.~Wang, X.~Yu, L.~Yang and Y.~Xu, \emph{Optical engineering}, 2017, \textbf{56}, 057105--057105\relax
\mciteBstWouldAddEndPuncttrue
\mciteSetBstMidEndSepPunct{\mcitedefaultmidpunct}
{\mcitedefaultendpunct}{\mcitedefaultseppunct}\relax
\EndOfBibitem
\bibitem[{G{\'o}mez-Medina} \emph{et~al.}(2011){G{\'o}mez-Medina}, {Garc{\'\i}a-C{\'a}mara}, {Su{\'a}rez-Lacalle}, {Gonz{\'a}lez}, {Moreno}, {Nieto-Vesperinas}, and {S{\'a}enz}]{GeDipole}
R.~{G{\'o}mez-Medina}, B.~{Garc{\'\i}a-C{\'a}mara}, I.~{Su{\'a}rez-Lacalle}, F.~{Gonz{\'a}lez}, F.~{Moreno}, M.~{Nieto-Vesperinas} and J.~J. {S{\'a}enz}, \emph{Journal of Nanophotonics}, 2011, \textbf{5}, 053512--053512\relax
\mciteBstWouldAddEndPuncttrue
\mciteSetBstMidEndSepPunct{\mcitedefaultmidpunct}
{\mcitedefaultendpunct}{\mcitedefaultseppunct}\relax
\EndOfBibitem
\bibitem[{\"O}zen(2020)]{2020SiSiO2DBR}
Y.~{\"O}zen, \emph{Applied Physics A}, 2020, \textbf{126}, 632\relax
\mciteBstWouldAddEndPuncttrue
\mciteSetBstMidEndSepPunct{\mcitedefaultmidpunct}
{\mcitedefaultendpunct}{\mcitedefaultseppunct}\relax
\EndOfBibitem
\bibitem[{\c{C}}etinkaya \emph{et~al.}(2022){\c{C}}etinkaya, {\c{C}}okduygulular, K{\i}nac{\i}, G{\"u}zel{\c{c}}imen, {\"O}zen, S{\"o}nmez, and {\"O}z{\c{c}}elik]{2022_1DPC}
{\c{C}}.~{\c{C}}etinkaya, E.~{\c{C}}okduygulular, B.~K{\i}nac{\i}, F.~G{\"u}zel{\c{c}}imen, Y.~{\"O}zen, N.~A. S{\"o}nmez and S.~{\"O}z{\c{c}}elik, \emph{Scientific Reports}, 2022, \textbf{12}, 11245\relax
\mciteBstWouldAddEndPuncttrue
\mciteSetBstMidEndSepPunct{\mcitedefaultmidpunct}
{\mcitedefaultendpunct}{\mcitedefaultseppunct}\relax
\EndOfBibitem
\bibitem[Vandana \emph{et~al.}(2024)Vandana, Ahmad, and Mallik]{2024MicroTexture}
L.~Vandana, G.~Ahmad and S.~Mallik, \emph{Materials Today: Proceedings}, 2024\relax
\mciteBstWouldAddEndPuncttrue
\mciteSetBstMidEndSepPunct{\mcitedefaultmidpunct}
{\mcitedefaultendpunct}{\mcitedefaultseppunct}\relax
\EndOfBibitem
\bibitem[Kumar \emph{et~al.}(2021)Kumar, Ramasesha,\emph{et~al.}]{2021NanoPillar}
D.~Kumar, S.~K. Ramasesha \emph{et~al.}, \emph{IEEE Transactions on Electron Devices}, 2021, \textbf{68}, 4504--4508\relax
\mciteBstWouldAddEndPuncttrue
\mciteSetBstMidEndSepPunct{\mcitedefaultmidpunct}
{\mcitedefaultendpunct}{\mcitedefaultseppunct}\relax
\EndOfBibitem
\bibitem[Sultan \emph{et~al.}(2024)Sultan, Al~Suny, Hossain, Noor, and Chowdhury]{2024NG}
R.~B. Sultan, A.~Al~Suny, M.~H. Hossain, T.~Noor and M.~H. Chowdhury, \emph{Heliyon}, 2024\relax
\mciteBstWouldAddEndPuncttrue
\mciteSetBstMidEndSepPunct{\mcitedefaultmidpunct}
{\mcitedefaultendpunct}{\mcitedefaultseppunct}\relax
\EndOfBibitem
\bibitem[Kazmi \emph{et~al.}(2020)Kazmi, Khan, Khan, Rauf, Farooq, Noman, and Ali]{2019DBR}
S.~A.~A. Kazmi, A.~D. Khan, A.~D. Khan, A.~Rauf, W.~Farooq, M.~Noman and H.~Ali, \emph{Applied Physics A}, 2020, \textbf{126}, 1--8\relax
\mciteBstWouldAddEndPuncttrue
\mciteSetBstMidEndSepPunct{\mcitedefaultmidpunct}
{\mcitedefaultendpunct}{\mcitedefaultseppunct}\relax
\EndOfBibitem
\bibitem[Ferdoushi \emph{et~al.}(2022)Ferdoushi, Wahid, and Alam]{2022FS}
M.~Ferdoushi, S.~Wahid and M.~K. Alam, \emph{RSC advances}, 2022, \textbf{12}, 19359--19374\relax
\mciteBstWouldAddEndPuncttrue
\mciteSetBstMidEndSepPunct{\mcitedefaultmidpunct}
{\mcitedefaultendpunct}{\mcitedefaultseppunct}\relax
\EndOfBibitem
\bibitem[Shen \emph{et~al.}(2023)Shen, Wang, Wangyang, and Zhou]{LigHarThin}
X.~Shen, Z.~Wang, P.~Wangyang and H.~Zhou, \emph{Optics Communications}, 2023, \textbf{545}, 129624\relax
\mciteBstWouldAddEndPuncttrue
\mciteSetBstMidEndSepPunct{\mcitedefaultmidpunct}
{\mcitedefaultendpunct}{\mcitedefaultseppunct}\relax
\EndOfBibitem
\bibitem[Mahmoud \emph{et~al.}(2019)Mahmoud, Hussein, Hameed, Abdel-Aziz, Hosny, and Obayya]{OptPerMod}
A.~H.~K. Mahmoud, M.~Hussein, M.~F.~O. Hameed, M.~Abdel-Aziz, H.~Hosny and S.~Obayya, \emph{JOSA B}, 2019, \textbf{36}, 357--365\relax
\mciteBstWouldAddEndPuncttrue
\mciteSetBstMidEndSepPunct{\mcitedefaultmidpunct}
{\mcitedefaultendpunct}{\mcitedefaultseppunct}\relax
\EndOfBibitem
\bibitem[Yu \emph{et~al.}(2017)Yu, Yao, Wu, Niu, Rogach, and Wang]{EffPlasMet}
P.~Yu, Y.~Yao, J.~Wu, X.~Niu, A.~L. Rogach and Z.~Wang, \emph{Scientific reports}, 2017, \textbf{7}, 7696\relax
\mciteBstWouldAddEndPuncttrue
\mciteSetBstMidEndSepPunct{\mcitedefaultmidpunct}
{\mcitedefaultendpunct}{\mcitedefaultseppunct}\relax
\EndOfBibitem
\bibitem[Eskandari and Habibzadeh-Sharif(2024)]{EnhLigAbsUltra}
M.~Eskandari and A.~Habibzadeh-Sharif, \emph{Photonics and Nanostructures-Fundamentals and Applications}, 2024, \textbf{58}, 101229\relax
\mciteBstWouldAddEndPuncttrue
\mciteSetBstMidEndSepPunct{\mcitedefaultmidpunct}
{\mcitedefaultendpunct}{\mcitedefaultseppunct}\relax
\EndOfBibitem
\bibitem[Zhu \emph{et~al.}(2010)Zhu, Hsu, Yu, Fan, and Cui]{NanodomeSolCel}
J.~Zhu, C.-M. Hsu, Z.~Yu, S.~Fan and Y.~Cui, \emph{Nano letters}, 2010, \textbf{10}, 1979--1984\relax
\mciteBstWouldAddEndPuncttrue
\mciteSetBstMidEndSepPunct{\mcitedefaultmidpunct}
{\mcitedefaultendpunct}{\mcitedefaultseppunct}\relax
\EndOfBibitem
\bibitem[Zhu \emph{et~al.}(2009)Zhu, Yu, Burkhard, Hsu, Connor, Xu, Wang, McGehee, Fan, and Cui]{OptAbsEnhAmor}
J.~Zhu, Z.~Yu, G.~F. Burkhard, C.-M. Hsu, S.~T. Connor, Y.~Xu, Q.~Wang, M.~McGehee, S.~Fan and Y.~Cui, \emph{Nano letters}, 2009, \textbf{9}, 279--282\relax
\mciteBstWouldAddEndPuncttrue
\mciteSetBstMidEndSepPunct{\mcitedefaultmidpunct}
{\mcitedefaultendpunct}{\mcitedefaultseppunct}\relax
\EndOfBibitem
\bibitem[Bosio \emph{et~al.}(2020)Bosio, Pasini, and Romeo]{History-of-Photovoltaics}
A.~Bosio, S.~Pasini and N.~Romeo, \emph{Coatings}, 2020, \textbf{10}, 344\relax
\mciteBstWouldAddEndPuncttrue
\mciteSetBstMidEndSepPunct{\mcitedefaultmidpunct}
{\mcitedefaultendpunct}{\mcitedefaultseppunct}\relax
\EndOfBibitem
\bibitem[Barbato \emph{et~al.}(2021)Barbato, Artegiani, Bertoncello, Meneghini, Trivellin, Mantoan, Romeo, Mura, Ortolani, Zanoni,\emph{et~al.}]{OperationandReliability}
M.~Barbato, E.~Artegiani, M.~Bertoncello, M.~Meneghini, N.~Trivellin, E.~Mantoan, A.~Romeo, G.~Mura, L.~Ortolani, E.~Zanoni \emph{et~al.}, \emph{Journal of Physics D: Applied Physics}, 2021, \textbf{54}, 333002\relax
\mciteBstWouldAddEndPuncttrue
\mciteSetBstMidEndSepPunct{\mcitedefaultmidpunct}
{\mcitedefaultendpunct}{\mcitedefaultseppunct}\relax
\EndOfBibitem
\bibitem[Tavakoli \emph{et~al.}(2016)Tavakoli, Lin, Leung, Lui, Lu, Li, Xiang, and Fan]{inano-perv}
M.~M. Tavakoli, Q.~Lin, S.-F. Leung, G.~C. Lui, H.~Lu, L.~Li, B.~Xiang and Z.~Fan, \emph{Nanoscale}, 2016, \textbf{8}, 4276--4283\relax
\mciteBstWouldAddEndPuncttrue
\mciteSetBstMidEndSepPunct{\mcitedefaultmidpunct}
{\mcitedefaultendpunct}{\mcitedefaultseppunct}\relax
\EndOfBibitem
\bibitem[Ding \emph{et~al.}(2020)Ding, Zhang, Mao, Xiao, Zhang, Wu, Zhang, and Jie]{img-patterned-Si}
K.~Ding, M.~Zhang, J.~Mao, P.~Xiao, X.~Zhang, D.~Wu, X.~Zhang and J.~Jie, \emph{Materials Today Energy}, 2020, \textbf{18}, 100493\relax
\mciteBstWouldAddEndPuncttrue
\mciteSetBstMidEndSepPunct{\mcitedefaultmidpunct}
{\mcitedefaultendpunct}{\mcitedefaultseppunct}\relax
\EndOfBibitem
\bibitem[Sakai \emph{et~al.}(2013)Sakai, Harada, Barada, and Fukuda]{chem-etch-slime}
D.~Sakai, K.~Harada, D.~Barada and T.~Fukuda, \emph{Japanese Journal of Applied Physics}, 2013, \textbf{52}, 036701\relax
\mciteBstWouldAddEndPuncttrue
\mciteSetBstMidEndSepPunct{\mcitedefaultmidpunct}
{\mcitedefaultendpunct}{\mcitedefaultseppunct}\relax
\EndOfBibitem
\bibitem[Ruiz-Ortega \emph{et~al.}(2023)Ruiz-Ortega, Esquivel-Mendez, Gonzalez-Trujillo, Hernandez-Vasquez, Matsumoto, and Albor-Aguilera]{AnalysisofCdS}
R.~C. Ruiz-Ortega, L.~A. Esquivel-Mendez, M.~A. Gonzalez-Trujillo, C.~Hernandez-Vasquez, Y.~Matsumoto and M.~d.~L. Albor-Aguilera, \emph{ACS omega}, 2023, \textbf{8}, 31725--31737\relax
\mciteBstWouldAddEndPuncttrue
\mciteSetBstMidEndSepPunct{\mcitedefaultmidpunct}
{\mcitedefaultendpunct}{\mcitedefaultseppunct}\relax
\EndOfBibitem
\bibitem[Santbergen \emph{et~al.}(2010)Santbergen, Liang, and Zeman]{silver-np-si}
R.~Santbergen, R.~Liang and M.~Zeman, 2010 35th IEEE Photovoltaic Specialists Conference, 2010, pp. 000748--000753\relax
\mciteBstWouldAddEndPuncttrue
\mciteSetBstMidEndSepPunct{\mcitedefaultmidpunct}
{\mcitedefaultendpunct}{\mcitedefaultseppunct}\relax
\EndOfBibitem
\bibitem[Shao \emph{et~al.}(2017)Shao, Chen, Guo, Zhang, Chang, Liu, Chen, Li, Li, and He]{organ-np}
P.~Shao, X.~Chen, X.~Guo, W.~Zhang, F.~Chang, Q.~Liu, Q.~Chen, J.~Li, Y.~Li and D.~He, \emph{Organic Electronics}, 2017, \textbf{50}, 77--81\relax
\mciteBstWouldAddEndPuncttrue
\mciteSetBstMidEndSepPunct{\mcitedefaultmidpunct}
{\mcitedefaultendpunct}{\mcitedefaultseppunct}\relax
\EndOfBibitem
\bibitem[Bollani \emph{et~al.}(2014)Bollani, Bietti, Frigeri, Chrastina, Reyes, Smereka, Millunchick, Vanacore, Burghammer, Tagliaferri,\emph{et~al.}]{orderedGa}
M.~Bollani, S.~Bietti, C.~Frigeri, D.~Chrastina, K.~Reyes, P.~Smereka, J.~Millunchick, G.~Vanacore, M.~Burghammer, A.~Tagliaferri \emph{et~al.}, \emph{Nanotechnology}, 2014, \textbf{25}, 205301\relax
\mciteBstWouldAddEndPuncttrue
\mciteSetBstMidEndSepPunct{\mcitedefaultmidpunct}
{\mcitedefaultendpunct}{\mcitedefaultseppunct}\relax
\EndOfBibitem
\bibitem[Aouassa \emph{et~al.}(2023)Aouassa, Bouabdellaoui, Yahyaoui, Kallel, Ettaghzouti, Algarni, and Althobaiti]{mngermaniumnp}
M.~Aouassa, M.~Bouabdellaoui, M.~Yahyaoui, T.~Kallel, T.~Ettaghzouti, S.~A. Algarni and I.~O. Althobaiti, \emph{ACS Applied Electronic Materials}, 2023, \textbf{5}, 2696--2703\relax
\mciteBstWouldAddEndPuncttrue
\mciteSetBstMidEndSepPunct{\mcitedefaultmidpunct}
{\mcitedefaultendpunct}{\mcitedefaultseppunct}\relax
\EndOfBibitem
\bibitem[Eze \emph{et~al.}(2021)Eze, Ugwuanyi, Li, Eze, Rodriguez, Evans, Rocha, Li, and Min]{bcsputter}
M.~C. Eze, G.~Ugwuanyi, M.~Li, H.~U. Eze, G.~M. Rodriguez, A.~Evans, V.~G. Rocha, Z.~Li and G.~Min, \emph{Solar Energy Materials and Solar Cells}, 2021, \textbf{230}, 111185\relax
\mciteBstWouldAddEndPuncttrue
\mciteSetBstMidEndSepPunct{\mcitedefaultmidpunct}
{\mcitedefaultendpunct}{\mcitedefaultseppunct}\relax
\EndOfBibitem
\end{mcitethebibliography}
\bibliographystyle{rsc} %the RSC's .bst file

\ifarXiv
    \foreach \x in {1,...,\numbersupplementpages}
    {
        \clearpage
        \includepdf[pages={\x},pagecommand={}]{\supplementfilename}
    }
\fi

\end{document}